\documentclass[journal]{IEEEtran}
\usepackage{graphicx, float}
\usepackage{placeins}
\usepackage{textcomp}
\usepackage{pgfplots}
\pgfplotsset{compat=1.18}
\usepackage{caption}
\usepackage{dblfloatfix}
\usepackage{hyperref}
\usepackage{algorithm}
\usepackage{algorithmicx}
\usepackage{algpseudocode}

\algnewcommand\algorithmicinput{\textbf{Input:}}
\algnewcommand\algorithmicoutput{\textbf{Output:}}
\algnewcommand\algorithmbegin{\textbf{BEGIN}}
\algnewcommand\BEGIN{\item[\algorithmbegin{}]}
\algnewcommand\Input{\item[\algorithmicinput{}]}
\algnewcommand\Output{\item[\algorithmicoutput{}]}
\algnewcommand\algorithmEnd{\textbf{END}}
\algnewcommand\END{\item[\algorithmEND{}]}
\algnewcommand\algorithmstepone{\textbf{Step 1:}}
\algnewcommand\Stepone{\item[\algorithmstepone{}]}
\algnewcommand\algorithmsteptwo{\textbf{Step 2:}}
\algnewcommand\Steptwo{\item[\algorithmsteptwo{}]}
\algnewcommand\algorithmstepthree{\textbf{Step 3:}}
\algnewcommand\Stepthree{\item[\algorithmstepthree{}]}
\usepackage{multirow}

\usepackage{lineno,hyperref}
\usepackage{colortbl}
\usepackage{rotating}
\usepackage{scalefnt}
\usepackage{tikz}

\usepackage{amsmath}
\usepackage{amssymb}
\usepackage{makecell}
\ifCLASSINFOpdf
\else
\fi
\begin{document}

\title{On the Effectiveness of Adversarial Samples against Ensemble Learning-based Windows PE Malware Detectors}
%
%
%

\author{\IEEEauthorblockN{Trong-Nghia To \IEEEauthorrefmark{1}\IEEEauthorrefmark{2},
Danh Le Kim\IEEEauthorrefmark{1}\IEEEauthorrefmark{2}, Do Thi Thu Hien\IEEEauthorrefmark{1}\IEEEauthorrefmark{2}, Nghi Hoang Khoa\IEEEauthorrefmark{1}\IEEEauthorrefmark{2}, \\Hien Do Hoang\IEEEauthorrefmark{1}\IEEEauthorrefmark{2}, Phan The Duy \IEEEauthorrefmark{1}\IEEEauthorrefmark{2}, and Van-Hau Pham\IEEEauthorrefmark{1}\IEEEauthorrefmark{2}}\\
\IEEEauthorblockA{\IEEEauthorrefmark{1}Information Security Laboratory, University of Information Technology, Ho Chi Minh city, Vietnam\\
\IEEEauthorrefmark{2}Vietnam National University, Ho Chi Minh city, Vietnam\\
nghiatt@uit.edu.vn, 18520560@gm.uit.edu.vn, \{hiendtt, khoanh, hiendh, duypt, haupv\}@uit.edu.vn}
 \thanks{Trong-Nghia To, Danh Le Kim, Do Thi Thu Hien, Nghi Hoang Khoa, Hien Do Hoang, Phan The Duy, Van-Hau Pham are with Information Security Lab (InSecLab), University of Information Technology, Vietnam National University Ho Chi Minh City, Hochiminh City, Vietnam. Website: (see at http://uit.edu.vn).}
\thanks{The corresponding author is Van-Hau Pham (Email: haupv@uit.edu.vn). }
\thanks{Manuscript received April 19, 20xx; revised August 26, 20xx.}
}

%
%

\markboth{Journal of \LaTeX\ Class Files,~Vol.~14, No.~8, August~20xx}%
{Shell \MakeLowercase{\textit{et al.}}: Bare Demo of IEEEtran.cls for IEEE Journals}
%



\maketitle
\pdfoutput=1
\begin{abstract}

Recently, there has been a growing focus and interest in applying machine learning (ML) to the field of cybersecurity, particularly in malware detection and prevention. Several research works on malware analysis have been proposed, offering promising results for both academic and practical applications. In these works, the use of Generative Adversarial Networks (GANs) or Reinforcement Learning (RL) can aid malware creators in crafting metamorphic malware that evades antivirus software. In this study, we propose a mutation system to counteract ensemble learning-based detectors by combining GANs and an RL model, overcoming the limitations of the MalGAN model. Our proposed FeaGAN model is built based on MalGAN by incorporating an RL model called the Deep Q-network anti-malware Engines Attacking Framework (DQEAF). The RL model addresses three key challenges in performing adversarial attacks on Windows Portable Executable malware, including format preservation, executability preservation, and maliciousness preservation. In the FeaGAN model, ensemble learning is utilized to enhance the malware detector's evasion ability, with the generated adversarial patterns. The experimental results demonstrate that 100\% of the selected mutant samples preserve the format of executable files, while certain successes in both executability preservation and maliciousness preservation are achieved, reaching a stable success rate.

\end{abstract}


%
\IEEEpeerreviewmaketitle

\pdfoutput=1
\section{Introduction}
%
%
%
%
\IEEEPARstart{W}{ITH} the advancement and rapid development of information technology, computer systems, and networks have become crucial and widespread in our daily lives. Besides that, cyberattacks are always a threat to cybersecurity that comes with development. Malicious Software (malware) is one of the most effective cyberattacks used by attackers to perform malicious behaviors such as stealing sensitive information without permission, affecting information systems, and demanding a massive ransom. In 2018, Symantec reported that 246,002,762 new malware variants emerged \cite{P1L_li2020adversarial}. Besides, among operating systems, Windows is the most wide-used one compared to other counterparts such as macOS, Android, Linux, etc. Hence, it has become the favorite target of attackers with malware in form of its PE (Portable Executable). According to Kaspersky Lab statistics at the end of 2020, an average of 360,000 malware are detected by Kaspersky every day, and more than 90\% of them are Windows PE \cite{P1S_ling2021adversarial} malware. 

To propose effective protection against malware threat, many researchers have applied ML and deep learning (DL) to malware detection. Those cutting-edge techniques have achieved success in various fields, as well as in feature extraction and classification of malware \cite{P3I_WindowsPEmalware}. However, ML and DL models are discovered to be vulnerable to adversarial attacks \cite{P1S_ling2021adversarial}, generated by slightly shuffling legitimate inputs, leading to judgment fallacy for the targeted models. Hence, considering the capability of dealing with adversarial samples when evaluating ML/DL-based solutions is a rising research trend. 

Many works have focused on the task of generating such adversarial samples based on modifying the original ones via various promising methods. Some of them intend to make modifications and create a complete adversarial sample, while others may just produce the adversarial form of representative data of malware, such as a feature vector. A possible approach is the Generative Adversarial Networks (GANs) model introduced by Goodfellow et al. \cite{11L_goodfellow2014GAN}. GANs has shown its potential in creating images, sounds, text and even in the field of information security to craft adversarial malware \cite{P24L_ImproveMalGAN,P26_GAPGAN}. However, using GANs still has limitations in generating adversarial samples when it only enables crafting adversarial features rather than executable malware samples.

\begin{figure*}[!b]
    \centering
    \includegraphics[width=0.75\textwidth]{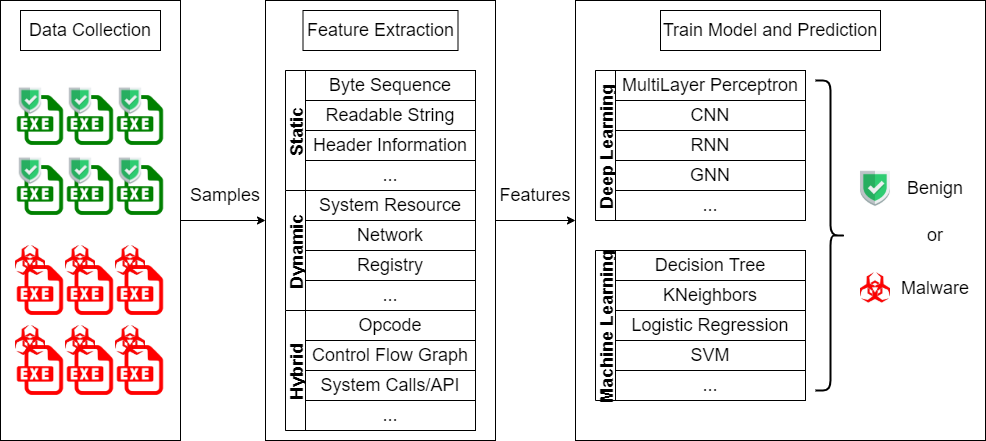}
    \caption{The general working flow of Malware Detection using ML/DL.}
    \label{fig:learning_framework}
\end{figure*}

Meanwhile, Reinforcement Learning (RL) is another potential solution for creating mutants of metamorphic malware \cite{P23L_gymmalware,P25L_DQEAF,AIMED-RL}. RL is a type of machine learning that involves an agent learning to interact with an environment through trial and error, receiving rewards for successful actions and punishments for unsuccessful ones. By using RL, it is possible to create malware that can adapt and evolve, making it more challenging for anti-malware systems to detect. On the other hand, GANs are primarily used for generating new data that resembles the training data, which may not be as effective in creating metamorphic malware. Though RL is still in the early stages of research in this area, it shows promise as a potential approach for creating more advanced and evasive malware. Moreover, while using GANs must be aware of the features to work on, RL can take action despite being totally unaware of this information but still produce modified samples.

Meanwhile, to deal with these adversarial attacks, more general and powerful methods are constantly emerging and evolving, one of which is Ensemble learning. This is a technique that combines several learning algorithms to increase the performance of the overall prediction \cite{2I_damavsevivcius2021ensemble}. Not only applied in malware defense, but this technique also gets more attention in the field of generating effective adversarial malware samples. According to Deqiang and Qianmu \cite{P1L_li2020adversarial}, there are two ensemble-based approaches to improve the effectiveness of adversarial samples: by using multiple attack methods and by attacking multiple classifiers. In the first approach, using multiple attack methods disturbs and increases the probability of misclassification in classifiers, such as the work of Tramèr et al. \cite{13L_tramer2019adversarial}. For the second approach, they use multiple classifiers to enable adversarial samples to interact as much as possible to increase the evasion of the samples towards them. Liu et al. \cite{14L_liu2016delving} proposed to improve the transferability of samples by attacking a group of combined DL models, rather than attacking a single model.


Motivated by the above promising solutions, our work aims to build a system to enhance the evasive effectiveness of Windows malware using the combination of GANs model and RL. Our proposed FeaGAN, inherited from the work of Hu and Tan \cite{1I_hu2017generating}, is designed with the ensemble learning method for training to take advantage of multiple models to generate adversarial features. Besides, we use RL to merge mutant vectors from FeaGAN into the original malicious PE files. This improves the evasion capability as well as enables verifying the executability and maliciousness of the malware.

The remainder of this work is organized as follows. In \textbf{Section~\ref{sec_background}}, we give the background of PE malware, ensemble learning, RL, as well as generative approaches of mutating malicious software. \textbf{Section~\ref{sec_methodology}} presents our method in crafting mutated malware against ensemble learning-based detectors by leveraging Reinforcement Learning (RL) and Generative Adversarial Networks (GAN). The experimental settings and results are given in \textbf{Section~\ref{sec_exp}}. \textbf{Section~\ref{sec_relatedwork}} discusses the related works of creating adversarial malware samples. Finally, in \textbf{Section~\ref{sec_conclusion}}, we conclude the paper and discuss future directions for this work.

\section{Background} \label{sec_background}

\subsection{ML-based Malware Detection}

Inspired from the success of ML/DL models in the fields of Computer Vision, Natural Language Processing (NLP), ML/DL-based malware detection methods have been proposed in PE malware detection. Due to their self-learning ability, ML/DL-based ones have a good generalization for unseen data. \textbf{Fig.~\ref{fig:learning_framework}} illustrates the required steps to create an ML/DL-based malware detector, including data collection, feature extraction, model learning from data, and predictions.

\subsubsection{Data Collection}

In fact, the quality of the data has a noticeable effect on the outcome of the ML/DL model \cite{28_Danh_cortes1994limits}. However, there is a lack of specific data standards for malware PE data like computer vision or NLP, while most cyber-security companies consider PE samples as their private property and rarely release them to the public \cite{P1S_ling2021adversarial}. Some researchers also publicly provide their PE files \cite{virusshare, 44_Danh_hastie2009elements, malwarebazaar}, but not in any specific standard. Moreover, the malware labeling task is mainly based on VirusTotal which uses a variety of antivirus tools to detect malware. However, in some cases, the variety of used tools can result in inconsistent results for the same sample. Hence, several methods have been proposed \cite{56_Danh_kantchelian2015better, 116_Danh_sebastian2016avclass, 117_Danh_sebastian2020avclass2, 152_Danh_zhu2020benchmarking} to unify the labeling based on the reliability of antivirus or voting. 

\begin{figure*}[!b]
    \centering
    \includegraphics[width=0.9\textwidth]{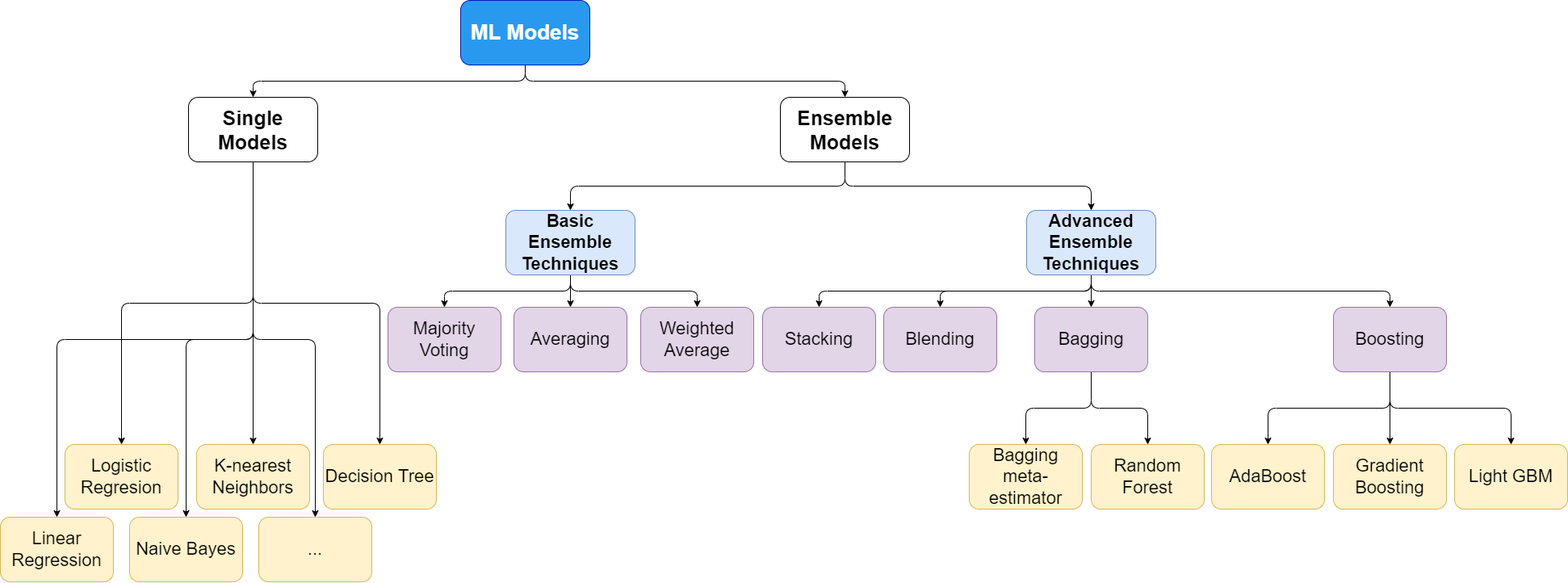}
    \caption{ML-models division diagram based on how it combines weak algorithms and complexity (excluding Single Models).}
    \label{fig:MLModel}
\end{figure*}

\subsubsection{Feature Extraction}

Once PE samples have the appropriate labels, it is necessary to extract useful features from those files and then transform them into a suitable format to use as input for ML/DL models. In fact, many ML/DL models accept numeric input, so those features are often converted into numbers. Useful features can help models gain knowledge to recognize malware and benign software. There are many features that can be obtained from a PE sample, which can be divided into three main categories: static, dynamic, and hybrid \cite{18_Danh_ceschin2020machine, 106_Danh_raff2020survey, 142_Danh_ye2017survey}, as shown in \textbf{Table~\ref{table:static_dynamic_hybrid}}. While static features are obtained directly from PE samples without running them, in some cases, the malware samples are executed in an isolated environment (sandbox, virtual machine) and all its behaviors that affect the environment are saved and extracted in terms of dynamic features. Meanwhile, hybrid features are extracted from the sample PE file by either static or dynamic approaches.

\begin{table}[t]
\centering
\caption{Some common types of features in PE files}
\begin{tabular}{cccc}
\hline
\multirow{2}{*}{\textbf{Features}} & \multicolumn{3}{c}{\textbf{Type}}           \\ \cline{2-4} 
                                    & \multicolumn{1}{c}{\textbf{Static}} & \textbf{Dynamic} & \textbf{Hybrid} \\ \hline
Byte Sequence                          & \multicolumn{1}{c}{x}             &              & \\ \hline
Readable Strings                      & \multicolumn{1}{c}{x}             &             &  \\ \hline
Header Information                  & \multicolumn{1}{c}{x}             &               &\\ \hline
System Resource Information    & \multicolumn{1}{c}{}              & x             &\\ \hline
File Information                       & \multicolumn{1}{c}{}              & x           &  \\ \hline
Registry Information                 & \multicolumn{1}{c}{}              & x            & \\ \hline
Network Information                      & \multicolumn{1}{c}{}              & x          &   \\ \hline
Opcode                &               &              & x             \\ \hline
System Calls/API              &               &              & x              \\ \hline
Control Flow Graph            &               &              & x             \\ \hline
Function Call Graph                     &               &              & x             \\ \hline
\end{tabular}
\label{table:static_dynamic_hybrid}
\end{table}

\subsubsection{Model training and Predictions}

After extracting features from the samples and converting them into numeric values, it is necessary to choose a suitable ML/DL model for malware and benign classification. Many ML/DL models have been proposed, such as Decision Tree (DT), Random Forest (RF), MultiLayer Perceptron (MLP), Naive Bayes, Support Vector Machine (SVM), Convolutional Neural Networks (CNN), Graph Neural Networks (GNN), Recurrent Neural Networks (RNN), or Long short-term memory (LSTM). They have achieved great success in the field of Computer Vision, NLP, and even vulnerability discovery \cite{81_Danh_ling2021deep, code_analysis_li2018vuldeepecker}. When it comes to malware detection, those models are also considered as potential solutions regardless of the diversity of features obtained from PE files, as long as the input matches the requirement of the model. The models attempt to learn how to recognize malware based on numerous training samples by seeking the relationships between features and their binary labels. The obtained knowledge from training can be used to predict the malware or benign label of unseen samples. The malware detection effectiveness can be different from model to model. Hence, many ML/DL models need to be considered and experimented to figure out the most suitable one for PE malware detection.

\subsection{Ensemble learning}

Ensemble learning, also known as multiple classifier systems or committee-based learning, aims to combine several base models to produce one optimal predictive model. The key idea of an ensemble learning method is to get benefits from different models by learning in an ensemble way. This can be a solution for cases of weak models or inconsistent results obtained from multiple models. Hence, putting them together in a suitable manner may result in a significantly better performance compared to using a single model. A simple form of the ensemble is to combine the results with majority voting.

 There are many methods to classify ensemble methods into many categories using different criteria. In the scope of this paper, we refer to the classification as depicted in \textbf{Fig.~\ref{fig:MLModel}}, where ensemble methods are categorized based on the combination mechanism and the complexity. This results in two groups of single models and ensemble models.

Many weaknesses of the single learning algorithm have motivated the development of ensemble methods. Most ensemble learning systems use learning models of the same type, which are called homogeneous ensembles. On the other hand, using different learning algorithms is called heterogeneous ensembles \cite{BookP3_zhou2012ensemble}. There are three main reasons which are statistical, computational, and representational \cite{17L_dietterich2000ensemble}. Learning algorithms try to get the best hypothesis in space. Because the amount of training data or training data is limited compared to the size of the hypothesis space, the statistical problem arises. This leads to a learning algorithm getting different hypotheses in space, which gives the same accuracy. The ensemble method helps this situation by averaging their votes and then reduces choosing the incorrect classifier and thus getting good accuracy on training data. Besides that, sometimes, the learning algorithms get stuck in the local optima even if we have enough training data. Using the ensemble method and then running, a local search from various origin points can lead to a better resemblance to the accurate unknown function as compared to the single base learner. In large cases of ML, it is hard to find a true function for the hypothesis space. By applying the various quantities of the hypothesis having weights, the space of representable functions can be expanded. One more reason is mentioned that ensemble methods are also so good when there is very little data as well as when there is too much \cite{bookP1_marsland2011machine}.


In an ensemble model, there are many factors that to be considered to create a reasonable model with good output. We focus on the following three approaches \cite{surveyP2_sagi2018ensemble}.
\subsubsection{Model training}
An ensemble model must consider two principles of diversity and predictive performance. While the diversity expects the participating inducers to be diverse enough to achieve the desired prediction performance through using various “inductive biases”, the predictive performance of each inducer should be as high as possible and should be at least as good as a random model. 
A model must have some inductive bias to be more useful when used with more data. The model's purpose is to fit most data, not just the sample data. As a result, inductive bias is critical. Furthermore, ensemble models with a variety of inducers may not always increase predictive performance. \cite{P18L_bi2012impact}.

\paragraph{Input manipulation} 
In this scenario, each base model is trained with a separate training subset, resulting in different inputs for the several base models. It is useful when tiny changes in the training set result in a different model.

\paragraph{Manipulated learning algorithm} 
In this approach, the use of each base model is changed. We can do this by modifying the way in which the base model traverses the hypothesis space.

\paragraph{Partitioning}
Diversity can be obtained by dividing a large dataset into smaller sub-sets and then using each subset to train various inducers.

\paragraph{Output manipulation}
This approach discusses the techniques that combine many binary classifiers into a single multi-class classifier. Error-correcting output codes (ECOC) are a successful example of this way.

\paragraph{Ensemble hybridization}
The idea is to combine at least two strategies when building the ensemble. The RF algorithm is probably the most well-known manifestation of the hybridization strategy. It not only manipulates the learning algorithm by selecting randomly a subset of features at each node, but also manipulates the instances when building each tree.
\subsubsection{Output fusion}
Output fusion discusses the process of merging the base model outputs into a single result. There are two main types we can refer to:

\paragraph{Weighting methods}
We can combine the base model outputs by assigning weights to each base model. The weighting method is most reasonable for instance where the performance of the base models is comparable. For classification problems, majority voting is the simplest weighting method. Another weighting strategy is to assign a weight that is proportional to the inducers’ strengths. 

\paragraph{Meta-learning methods}
Meta-learning models are different from standard ML models in that they involve more than one learning stage. In the meta-learning model, the individual inducer outputs are used as input to the meta-learner, which creates the final output. Meta-learning methods work well in cases where certain base models have different performances on various sub-spaces. Stacking is probably the most popular meta-learning method.
\subsubsection{Framework}
We can divide it into two main types of, the dependent framework and the independent one.

Regarding the dependent framework, the output of each inducer influences the construction of the next inducer. In this framework, information from the previous iteration guides learning in the next iteration. On the other hand, each inducer in the independent framework is independently built with other inducers. \textbf{Table~\ref{table:Ensemblemethodcategories}} shows the types of ensemble methods based on the above approaches.

\begin{table}[!t]
\centering
\caption{Ensemble method categories}
\begin{tabular}{lccc}
\hline
\multicolumn{1}{c}{\cellcolor[HTML]{FFFFFF}\textbf{Method name}} & \textbf{Fusion method} & \textbf{Dependency} & \textbf{\makecell{Training\\approach}} \\
\hline
Stacking                   & Meta-learning & Independent & \makecell[l]{Manipulated\\learning}   \\
\hline
AdaBoost                   & Weighting     & Dependent   & \makecell[l]{Input\\manipulation}   \\
\hline
\makecell[l]{Gradient Boosting\\(GB) machines} & Weighting     & Dependent   & \makecell[l]{Output\\manipulation}    \\
\hline
Random Forest              & Weighting     & Independent & \makecell[l]{Ensemble\\hybridization} \\
\hline
Bagging                    & Weighting     & Independent & \makecell[l]{Input\\manipulation}     \\
\hline
\end{tabular}%

\label{table:Ensemblemethodcategories}
\end{table}

\subsection{Generative Adversarial Networks and Reinforcement Learning}
Generative Adversarial Networks (GAN) is a technique for both semi-supervised and unsupervised learning, proposed by Goodfellow \cite{11L_goodfellow2014GAN}. The GAN architecture consists of two main parts including a generative model \textit{G} and a discriminative model \textit{D}. While \textit{G} tries to generate fake data with the aim of making it resemble the real one, \textit{D} is responsible for distinguishing between real samples and generated samples from \textit{G}. Both networks are trained concurrently and compete in a minimax two-player game.

When it comes to Reinforcement Learning (RL), it is also a branch of ML, besides supervised learning and unsupervised learning. This learning method relies on an \textit{agent} deciding to perform a suitable \textit{action} and then interacting with \textit{the environment} to get the best \textit{reward} in a specific state and get a new \textit{state} in return. More specifically, RL is distinct from supervised learning in that the training data is not pre-labeled, but rather learned through a process of trial and error. The RL model repeatedly attempts to solve the problem at hand and learns from the outcome of each attempt to gradually develop a suitable strategy. By contrast, supervised learning relies on pre-labeled training data to train the model.

\section{Methodology} \label{sec_methodology}

\begin{figure*}[!b]
    \centering
    \includegraphics[width=0.8\textwidth]{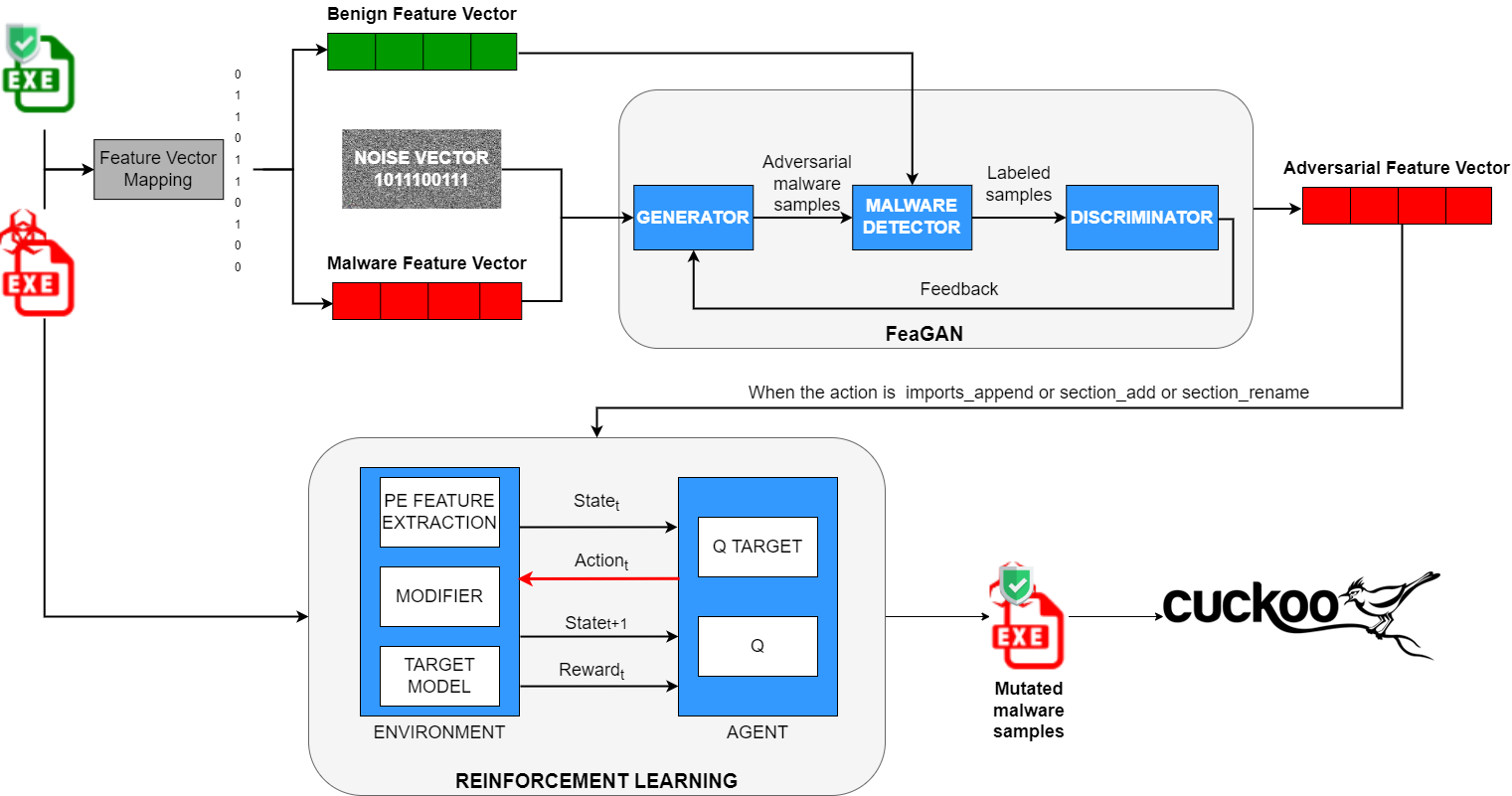}
    \caption{Overview of adversarial malware generation system.}
    \label{fig:methodology}
\end{figure*}

\subsection{Threat models}
Based on the knowledge of the target that got in the attacker's hand when performing the attack, attack scenarios can be classified as white-box, gray-box, and black-box. In white-box scenarios, it is assumed that the attacker has complete knowledge of the target including its training data, algorithms, and ML models along with training hyperparameters. Whereas in the gray-box scenarios, the attacker can only obtain limited or partial of that information. Opposite to white-box one, the attacker in the case of black-box is entirely oblivious to the target ML system without any knowledge. In fact, there are many opinions that it is impossible to perform an actual black-box attack. The reason is that the attacker must gather at least certain information, such as the location of the target ML model or its output, corresponding to a specific provided data. In the malware domain, a black-box attack typically refers to an attack on a target model where the attacker has only access to the target's input and output interfaces. 

Our threat model is defined in five following aspects.
\begin{itemize}
    \item \textbf{\textit{Knowledge of attacker:}} the obtained information about the target model of attacker. Our proposed method performs attacks in a black-box scenario. Mean that, the parameters and ensemble learning architecture of the malware detectors are not available to the attacker. Moreover, the attacker does not have access to any confidence score from the detectors. The only accessible information is whether the mutant samples can evade detection.
    \item \textbf{\textit{Manipulation Space:}} the nature of creating adversarial samples, which can be customized in problem space or feature space. Our method aims to work on both those spaces. We apply GAN to create adversarial feature vectors based on the original ones (feature space). Then, those created vectors can be leveraged during the modification process to produce actual malware samples by RL (problem space).
    \item \textbf{\textit{Attack Strategy:}} type of evasion attack. In our work, we utilize mimicry attack \cite{mimicryattack}, an evasion technique that seeks to transfer the attack point into a benign area, or attempts to mimic a specified benign point.
    \item \textbf{\textit{Target model:}} the detection model we focus to evade in the scope of this paper, which is Ensemble Learning-based detectors. 
    \item \textbf{\textit{Attacker's Goal:}} breaking the integrity of the C-I-A (Confidentiality - Integrity - Availability) model and creating new feature vectors as well as malware samples that can fool the Ensemble Learning-based detectors.
\end{itemize}

Besides, we also consider the transferability of adversarial attacks when evaluating the ability of mutant patterns generated during interaction with a given model to fool other ML models. The models in our study are single models and ensemble learning models.



\subsection{Our strategy}
Our proposed method to generate adversarial samples is made of two main parts of FeaGAN and the RL model, as shown in \textbf{Fig.~\ref{fig:methodology}}. In this system, FeaGAN takes the responsibility of generating adversarial feature vectors. Our FeaGAN is inspired by MalGAN \cite{1I_hu2017generating} with some improvements in the malware detection. 
This component only performs its duty on the feature vectors or feature space, without directly generating a complete malware sample. In this case, the RL model can be used to overcome this limitation when utilizing FeaGAN independently. By using the RL model, it helps us to decide the sequence of modifications to perform. If the agent chooses the actions to modify such as adding a new section, an import function, or changing a section name, the adversarial feature vectors generated from FeaGAN will be used. The actions of RL agent are all operations to transform a malicious software file without breaking its format.

\subsubsection{FeaGAN model}
FeaGAN incorporates three essential components including the Generator, malware detector, and Discriminator. Our proposed model takes advantage of the API call list to generate deceptive PE malware samples. The underlying assumption of FeaGAN is that attackers have complete knowledge of the feature space of the target malware detector. Consequently, the model seeks to construct a Discriminator that closely emulates the target malware detector, while concurrently training the Generator. The training process strengthens the generated adversarial malware by introducing spurious API calls to the original malware sample, which helps to deceive the Discriminator more efficiently.

\begin{itemize}
    \item \textbf{Generator:}
 In general, the Generator, which is indeed a multi-layer feed-forward neural network, is used to transform a malware feature vector \textit{m} into its adversarial version by adding noise \textit{z}. Each element of \textit{m} represents the presence or absence of a feature, while noise \textit{z} is a 10-dimension vector randomized in the range [0, 1]. The Generator is designed with two hidden layers of 256 nodes and Leaky ReLU as the activation function (\textbf{Eq.~(\ref{equation:leakyrelu})}). The output layer has 11,041 nodes for the corresponding number of features in the Malware Dataset. All these nodes have their output ensured in the range of (0,1) by the Sigmoid activation function. Moreover, we resolve the \textit{exploding gradient} problem by bounding the output of the Generator within the range $[\epsilon, 1 - \epsilon]$ with $\epsilon=10^{-7}$. 
 
    \begin{equation}
        g(x)=
        \begin{cases}
            x & x \geq 0\\
            \alpha x & $Otherwise$
        \end{cases}
        \label{equation:leakyrelu}
    \end{equation}
    \item \textbf{Malware detector:}
It plays a role as the third-party malware detector to label each input sample as either malware or benign. To implement this element, we use both single algorithms and ensemble learning ones to enable adversarial samples to interact with multiple classifiers to improve their evasiveness \cite{P1L_li2020adversarial}. In the case of ensemble learning, the Stacking algorithm is chosen due to its popularity and the benefits gained from each base model. Its pseudocode is given in \textbf{Algorithm~\ref{algo:Stacking}}. 

    \item \textbf{Discriminator:}
The Discriminator is utilized to mimic the operations of the malware detector and supplies gradient information to train the Generator. It is also a multi-layer feed-forward neural network with the same structure as Generator. The main difference is the output layer, where Discriminator has only 1 node. It also uses the Sigmoid activation function to show the \textit{probability} of the input vector to be malware.
\end{itemize}

\begin{algorithm}[!t]

\caption{Pseudocode of Stacking algorithm for malware detection}
\label{algo:Stacking}
\begin{algorithmic}[1]
\Input{ Training dataset $\mathcal{D} = \{ (x_1,y_1),(x_2,y_2),...,(x_n,y_n) \}$

Base level classifiers $\mathcal{C}_1,\mathcal{C}_2,...,\mathcal{C}_k$

Meta level classifier
$\hat{\mathcal{C}}$}
\Output{Trained ensemble classifier $\hat{\mathcal{S}}$}

\Stepone{Train base learners by applying classifiers $\mathcal{C}_i$ to dataset $\mathcal{D}$}
\For{\texttt{$i = 1, 2,..., k$}}
    \State \texttt{$\mathcal{B}_i = \mathcal{C}_i (\mathcal{D}) $}
\EndFor
\Steptwo{Construct new dataset of predictions $\hat{\mathcal{D}}$}
\For{\texttt{$i = 1, 2,..., n$}}
    \For{\texttt{$j = 1, 2,..., k$}}
        \State \textit{\# Use $\mathcal{B}_j$ to classify training example $x_i$}
        \State \texttt{$z_{ij} = \mathcal{B}_j(x_i)$}     
    \EndFor
    \State \texttt{$\hat{\mathcal{D}} = \{Z_i, y_i\}$}, where \texttt{$Z_i = \{z_{i1}, z_{i2},..., z_{ik}\}$}
\EndFor
\Stepthree{Train a meta-level classifier $\hat{\mathcal{M}}$}
\State \texttt{$\hat{\mathcal{M}} = \hat{\mathcal{C}} (\hat{\mathcal{D}}) $}
\State \Return $\hat{\mathcal{M}}$

\end{algorithmic}
\end{algorithm}

\subsubsection{RL model}\label{DQEAF}
We utilize the DQEAF \cite{P25L_DQEAF} as an RL model to assist the FeaGAN to generate complete adversarial malware samples instead of adversarial vectors. Given the reward and the state of the environment as the input, an action is chosen based on a rational strategy for the agent to perform in the next round. Additionally, each component is described as a Markov Decision Process (MDP) model to apply DQEAF to this problem:
\begin{itemize}
    \item State $s_t$ is a feature vector of the malware PE file.
    \item Given a state $s_t$, the agent will observe the state of the environment and choose an action $a_t$ from action space $A$ of available actions. In our RL model, we implement 10 actions defined as in \textbf{Table~\ref{tab-rl-action}}.
\begin{table}[!t]
    \centering
    \caption{Actions in action space of RL model}
    \begin{tabular}{cp{0.28\textwidth}}
    \hline
         \textbf{Action} & \textbf{Description} \\
    \hline
     overlay\_append    &  Randomly add some bytes to the end of the file \\
     \hline
     imports\_append & Randomly add an import function that is not in the file \\
     \hline
     section\_rename & Rename a section contained in the file \\
     \hline
     section\_add & Randomly add a section to the file\\
     \hline  
     section\_append & Randomly add some bytes to a section\\
     \hline
     upx\_pack & Pack files using UPX\\
     \hline
     upx\_unpack & Unpack the packaged executable using UPX\\
     \hline
     remove\_signature & Remove the signed certificate for the executable\\
     \hline
     remove\_debug & Remove debugging information in the file\\
     \hline
     break\_header\_checksum & Modify (break) header checksum\\
     \hline
    \end{tabular}
    
    \label{tab-rl-action}
\end{table}
    \item Reward $r_t$ is determined for each training $TURN$ based on the label from the detector and the number of performed actions. Moreover, $MAXTURN$ indicates that the agent should claim failure if $MAXTURN$ steps of modification have been taken. Reward $r_t$ is 0 if the label is malware, and is calculated using \textbf{Eq. (\ref{equation:reward_t})} when the label is benign.
    \begin{equation}
    \label{equation:reward_t}
        r_t = 20^{-(TURN-1)/MAXTURN}*100
    \end{equation}
    
    The agent is trained to maximize the expectation of the cumulative discounted reward given by \textbf{Eq. (\ref{equation:R})}.
    \begin{equation}
    \label{equation:R}
        R = \sum_{t=1}^T \gamma^{t-1} r_t
    \end{equation}
    
    where $\gamma \in [0,1]$ is a factor discounting future rewards.
    \item 
    The optimal action-value function $\hat{Q}$ which shown in \textbf{Eq. (\ref{equation:optimal_action_value})} estimates the expected reward of taking an action $a$ at state $s_t$.
    \begin{equation}
    \label{equation:optimal_action_value}
        \hat{Q} (s_t,a_t) = E\{ r_{t}+\gamma      \mathop{max}_{a_{t+1}} \hat{Q}(s_{t+1},a_{t+1})\}
    \end{equation}
\end{itemize}
The agent optimizes the weight $\theta$ to minimize the error estimated by the loss function (\textbf{Eq. (\ref{equation:loss})}) during learning process.
\begin{equation}
\label{equation:loss}
\begin{aligned}
        loss_t (\theta_t) = [(r+\gamma\mathop{max}_{a_{t+1}} Q(s_{t+1},a_{t+1};\theta_{t-1})) \\
        - Q(s_t,a_t;\theta_t)]^2
\end{aligned}
\end{equation}

\subsubsection{Cuckoo-based functional testing environment}
The functional testing environment is an isolated place from the system to capture and verify suspected malware. It is used to prevent these files from executing undesired behaviors affecting on the real system. We use the Cuckoo Sandbox, an open-source dynamic analysis system, as a testing environment for functional validation to evaluate the executability and maliciousness of created malwares.

\section{Experiments and Analysis} \label{sec_exp}

\subsection{Environment implementation}
Our system is implemented on the Ubuntu 18.04 LTS virtual machine, with a detailed hardware configuration of 16-core Intel Xeon E5-2660 CPU clocked at 2.0 GHz with 16 GB of RAM and a 300 GB disk. The source code is written in Python using the main libraries such as Pytorch, Scikit-learn, LIEF, and some other support libraries. It should be mentioned that the Stacking approach requires Scikit-Learn library version 0.22 or higher, which is the library that we install and utilize ensemble strategies.
\subsubsection{FeaGAN module}
The FeaGAN model is trained to generate new features for actions in the RL model, such as imports\_append or section\_add or section\_rename. The Generator and the Discriminator are both multi-layer feed-forward neural networks consisting of 2 hidden layers with 256 nodes in each layer. The activation function is Leaky ReLU defined in \textbf{Eq.~(\ref{equation:leakyrelu})} with $\alpha = 0.01$. The output layer has 11,041 nodes including 9,890 imports and 1,151 sections. This layer also uses a sigmoid activation function to ensure the output lies in the range (0, 1). The FeaGAN model is trained with 100 epochs, and a batch size of 32.
\subsubsection{RL module}
We implement the same techniques as described in Fang's DQEAF framework \cite{P25L_DQEAF}, but with 2,350 dimensions as in the gym-malware \cite{P23L_gymmalware}, for a more comprehensive view of malware. The RL agent has 10 possible actions, as listed in \textbf{Table \ref{tab-rl-action}} and is trained in a Deep Convolutional Q-network with two hidden layers with 256 and 64 nodes respectively. The activation function used in both layers is the ReLU function.

The agent training gets through over 600 episodes with a discount factor $\gamma$ of 0.99. In each episode, the agent is allowed to perform up to 80 actions on each PE file. If it receives a reward of 10 before reaching that limitation, it proceeds to the next episode to learn in a new state.

\subsubsection{FeaGAN's malware detectors}
\label{single_ensemble_define}
Note that, due to the high dimensions of the dataset \cite{high_dimension_wojcik2019training} described in \textbf{Section \ref{preprocess_FeaGAN}}, we carefully choose suitable algorithms that can adapt to that requirement. To implement single learning-based detectors, we utilize 5 algorithms of Decision Tree (DT), Logistic Regression (LR), Kneighbors (KNN), Naive Bayes and Bernoulli. Meanwhile, ensemble algorithms are deployed in homogeneous and heterogeneous manner. More specifically, Random Forest, Bagging, AdaBoost, Gradient Boosting are used as homogeneous algorithms. Besides, in the Voting technique, all five above single algorithms work as estimators with soft voting for prediction. In the Stacking technique, there are 2 different implementations. The first one has base estimators consisting of the three best single algorithms. The second one employs all five single algorithms as base estimators. From the group of the top three single methods indicated above, the final estimator is chosen.

\subsubsection{Target RL models}

To deploy target RL models, most of the other works employ a GB-based model and the gym-malware framework \cite{P23L_gymmalware,P25L_DQEAF}. In this paper, we also implement our GB-based target model, and then comparing with the GB-based one in gym-malware and other ensemble algorithms in the homogeneous approach as same as FeaGAN. Moreover, 5 algorithms of LR, DT, Naive Bayes, KNN and MLP are used as single models. Besides, in the heterogeneous approach, voting also uses all these five single algorithms. In the stacking method, we also have 2 cases of base estimators. The first one takes all five single algorithms as base estimators. Another one has base estimators of DT, RF, Adaboost, Bagging, GB. 

\subsection{Dataset}

\begin{table}[!b]
\caption{Overview of data distribution}
\centering
\resizebox{\linewidth}{!}{%
\begin{tabular}{ccccc}
\hline
\multicolumn{3}{c}{\multirow{2}{*}{\textbf{Dataset}}}                            & \multicolumn{2}{c}{\textbf{Label}}                     \\ \cline{4-5} 
\multicolumn{3}{c}{}                                                             & \multicolumn{1}{c}{\textbf{Benign}} & \textbf{Malware} \\ \hline
\multicolumn{1}{c}{\multirow{4}{*}{Dataset for FeaGAN}} & \multicolumn{1}{c}{\multirow{2}{*}{Malware detector}} & Training & \multicolumn{1}{c}{4,000}  & 4,800  \\ \cline{3-5} 
\multicolumn{1}{c}{} & \multicolumn{1}{c}{}                          & Testing  & \multicolumn{1}{c}{1,000}            & 1,600             \\ \cline{2-5} 
\multicolumn{1}{c}{} & \multicolumn{1}{c}{\multirow{2}{*}{GAN}}      & Training & \multicolumn{1}{c}{0}               & 1,600             \\ \cline{3-5} 
\multicolumn{1}{c}{} & \multicolumn{1}{c}{}                          & Testing  & \multicolumn{1}{c}{0}               & 1,600             \\ \hline
\multicolumn{1}{c}{\multirow{4}{*}{Dataset for RL}}     & \multicolumn{1}{c}{\multirow{2}{*}{Target model}}     & Training & \multicolumn{1}{c}{40,000} & 40,000 \\ \cline{3-5} 
\multicolumn{1}{c}{} & \multicolumn{1}{c}{}                          & Testing  & \multicolumn{1}{c}{10,000}           & 10,000            \\ \cline{2-5} 
\multicolumn{1}{c}{} & \multicolumn{1}{c}{\multirow{2}{*}{RL agent}} & Training & \multicolumn{1}{c}{0}               & 1,600             \\ \cline{3-5} 
\multicolumn{1}{c}{} & \multicolumn{1}{c}{}                          & Testing  & \multicolumn{1}{c}{0}               & 1,600             \\ \hline
\multicolumn{2}{c}{\textbf{Testset for the proposed system}}          & Testing  & \multicolumn{1}{c}{0}               & 2,000             \\ \hline
\end{tabular}%
}
\label{table:data_distribution}
\end{table}
\label{preprocess_FeaGAN}
Our dataset has 115,000 PE files extracted from \cite{dataset}, including 55,000 benign files collected from Windows 7 virtual machine (VM) and 60,000 malware files from VirusTotal, with labels of Adware, Trojan, Virus, Ransomware, Backdoor.

From the given dataset, features including the import functions and the sections are extracted. More specifics,  import functions are stored in a set in the format of $<\!\!\!\!\!Function\!\!\!\!\!>:<\!\!\!\!Name~of~DLL~library\!\!\!\!>$, for example, $ReadFile\!\!:\!kernel32.dll$, and then concatenated with Section Name (.text, .data, .bss, etc.). All features are represented as binary ones, which is 1 when the feature exists in the sample, and 0s for vice versa. 

This dataset is divided into different parts for purposes of training or testing in various components in our model, as described in \textbf{Table~\ref{table:data_distribution}}. In more detail, 5,000 benign files and 8,000 malware files are used in training and testing FeaGAN with different ratio. 80\% of benign samples take part in the training process of detector, while the remain benign ones are used in testing of this component. Meanwhile, malware samples are split up to 60\% and 20\% for training and testing the detector respectively, when the other 20\% plays its role in training GAN. We reuse the detector's testing data for the testing process in GAN.



In the RL model, we reuse the GAN dataset to train and test the RL agent (i.e., 3,200 malicious samples in a 50:50 ratio for training and testing). We also utilize 50,000 benign files and 50,000 malware files to initially train and test target models with a ratio of 8:2. 

After finishing the training components, we test our proposed system on the remaining 2,000 malicious samples against the target models.

\begin{table*}[!t]
\caption{Performance of Malware Detector in FeaGAN}
\centering
\begin{tabular}{ccccccc}
\hline
\multicolumn{2}{c}{\textbf{Algorithms}}                                                                    & \multicolumn{1}{c}{\textbf{AUC}}     & \multicolumn{1}{c}{\textbf{Accuracy}} & \multicolumn{1}{c}{\textbf{Precision}} & \multicolumn{1}{c}{\textbf{Recall}}  & \textbf{F1-score} \\ \hline
\multicolumn{1}{c}{\multirow{5}{*}{\textbf{Single}}}        & \multicolumn{1}{c}{Bernoulli}               & \multicolumn{1}{c}{0.84906} & \multicolumn{1}{c}{0.73115}  & \multicolumn{1}{c}{0.85839}   & \multicolumn{1}{c}{0.67438} & 0.75534  \\ \cline{2-7} 
\multicolumn{1}{c}{}                               & \multicolumn{1}{c}{Naive Bayes}             & \multicolumn{1}{c}{0.61137} & \multicolumn{1}{c}{0.69615}  & \multicolumn{1}{c}{0.67442}   & \multicolumn{1}{c}{\cellcolor{blue!25}\textbf{0.97875}} & 0.79857  \\ \cline{2-7} 
\multicolumn{1}{c}{}                               & \multicolumn{1}{c}{DT}            & \multicolumn{1}{c}{0.91580}  & \multicolumn{1}{c}{0.92231}  & \multicolumn{1}{c}{0.95155}   & \multicolumn{1}{c}{0.92063} & 0.93583  \\ \cline{2-7} 
\multicolumn{1}{c}{}                               & \multicolumn{1}{c}{LR}      & \multicolumn{1}{c}{\cellcolor{blue!25}\textbf{0.96459}} & \multicolumn{1}{c}{\cellcolor{blue!25}\textbf{0.92769}}  & \multicolumn{1}{c}{\cellcolor{blue!25}\textbf{0.95315}}   & \multicolumn{1}{c}{0.92813} & \cellcolor{blue!25}\textbf{0.94047}  \\ \cline{2-7} 
\multicolumn{1}{c}{}                               & \multicolumn{1}{c}{KNN}              & \multicolumn{1}{c}{0.94977} & \multicolumn{1}{c}{0.90231}  & \multicolumn{1}{c}{0.94335}   & \multicolumn{1}{c}{0.89500}   & 0.91854  \\ \hline
\multicolumn{1}{c}{\multirow{4}{*}{\textbf{Homogeneous}}}   & \multicolumn{1}{c}{RF}            & \multicolumn{1}{c}{\cellcolor{blue!25}\textbf{0.97978}} & \multicolumn{1}{c}{\cellcolor{blue!25}\textbf{0.93654}}  & \multicolumn{1}{c}{0.96320}    & \multicolumn{1}{c}{0.93250}  & \cellcolor{blue!25}\textbf{0.94760}   \\ \cline{2-7} 
\multicolumn{1}{c}{}                               & \multicolumn{1}{c}{Bagging}         & \multicolumn{1}{c}{0.97268} & \multicolumn{1}{c}{0.93115}  & \multicolumn{1}{c}{\cellcolor{blue!25}\textbf{0.96468}}   & \multicolumn{1}{c}{0.92188} & 0.94279  \\ \cline{2-7} 
\multicolumn{1}{c}{}                               & \multicolumn{1}{c}{AdaBoost}                & \multicolumn{1}{c}{0.95289} & \multicolumn{1}{c}{0.88500}    & \multicolumn{1}{c}{0.91249}   & \multicolumn{1}{c}{0.89938} & 0.90589  \\ \cline{2-7} 
\multicolumn{1}{c}{}                               & \multicolumn{1}{c}{GB}        & \multicolumn{1}{c}{0.96610}  & \multicolumn{1}{c}{0.92423}  & \multicolumn{1}{c}{0.93816}   & \multicolumn{1}{c}{\cellcolor{blue!25}\textbf{0.93875}} & 0.93846  \\ \hline
\multicolumn{1}{c}{\multirow{7}{*}{\textbf{Heterogeneous}}} & \multicolumn{1}{c}{Voting}                  & \multicolumn{1}{c}{0.96969} & \multicolumn{1}{c}{0.93000}     & \multicolumn{1}{c}{0.94479}   & \multicolumn{1}{c}{0.94125} & 0.94302  \\ \cline{2-7} 
\multicolumn{1}{c}{}                               & \multicolumn{1}{c}{Stacking(3, LR)}     & \multicolumn{1}{c}{\cellcolor{blue!25}\textbf{0.97738}} & \multicolumn{1}{c}{0.93615}  & \multicolumn{1}{c}{0.96021}   & \multicolumn{1}{c}{0.93500}   & 0.94594  \\ \cline{2-7} 
\multicolumn{1}{c}{}                               & \multicolumn{1}{c}{Stacking(3, KNN)}   & \multicolumn{1}{c}{0.95943} & \multicolumn{1}{c}{0.93192}  & \multicolumn{1}{c}{\cellcolor{blue!25}\textbf{0.96231}}   & \multicolumn{1}{c}{0.92563} & 0.94361  \\ \cline{2-7} 
\multicolumn{1}{c}{}                               & \multicolumn{1}{c}{Stacking(3, DT)} & \multicolumn{1}{c}{0.90274} & \multicolumn{1}{c}{0.91192}  & \multicolumn{1}{c}{0.94484}   & \multicolumn{1}{c}{0.91000}    & 0.92710   \\ \cline{2-7} 
\multicolumn{1}{c}{}                               & \multicolumn{1}{c}{Stacking(5, LR)}      & \multicolumn{1}{c}{0.97576} & \multicolumn{1}{c}{\cellcolor{blue!25}\textbf{0.93846}}  & \multicolumn{1}{c}{0.96038}   & \multicolumn{1}{c}{0.93875} & \cellcolor{blue!25}\textbf{0.94943}  \\ \cline{2-7} 
\multicolumn{1}{c}{}                               & \multicolumn{1}{c}{Stacking(5, KNN)}    & \multicolumn{1}{c}{0.74160}  & \multicolumn{1}{c}{0.78846}  & \multicolumn{1}{c}{0.76382}   & \multicolumn{1}{c}{\cellcolor{blue!25}\textbf{0.95000}}    & 0.84680   \\ \cline{2-7} 
\multicolumn{1}{c}{}                               & \multicolumn{1}{c}{Stacking(5, DT)}  & \multicolumn{1}{c}{0.90568} & \multicolumn{1}{c}{0.91538}  & \multicolumn{1}{c}{0.94344}   & \multicolumn{1}{c}{0.91750}  & 0.93029  \\ \hline
\end{tabular}%
\label{table:malware_FeaGAN_performance}
\end{table*}

\subsection{Performance Metrics}
The malware recognition performance of the detector is represented via metrics including Area Under Curve (AUC), accuracy, precision, recall, and F1-score. The above metrics are calculated based on the attributes of the Confusion Matrix, which are True Positive (TP), True Negative (TN), False Positive (FP), and False Negative (FN). In security, where positive and negative represent malicious and benign samples, respectively, those attributes are defined as follows.

\begin{itemize}
    \item TP: quantity of malware samples are truly classified.
    \item TN: quantity of benign samples are true classified.
    \item FN: quantity of malware samples are classified as benign.
    \item FP: quantity of benign samples are classified as malware.
\end{itemize}

Based on those definitions, \textit{Accuracy} is the ratio of correct predictions $TP,~TN$ in the total resulted ones, as in \textbf{Eq.~(\ref{eq:acc})}. Meanwhile, \textit{Precision}, as in \textbf{Eq.~(\ref{eq:pre})}, is the ratio of accurate malware $TP$ to the total number of malware labeled by the detector. Another metric called \textit{Recall}, which is defined in \textbf{Eq.~(\ref{eq:recall})}, measures the proportion of $TP$ over all malware instances in the testing dataset. \textit{F1-Score} reflects $Precision$ and $Recall$ in a single formula, as in \textbf{Eq.~(\ref{eq:f1})}. Besides, AUC is a metric to measure the quality of the virtualization of the tradeoff between TP rate and FP rate. Higher values in the above metrics indicate a more powerful malware detector. 

\begin{equation}\label{eq:acc}
Accuracy = \frac{TP + TN}{TP + TN + FP + FN}
\end{equation}

\begin{equation}\label{eq:pre}
Precision = \frac{TP}{TP + FP}
\end{equation}

\begin{equation}\label{eq:recall}
Recall = \frac{TP}{TP + FN}
\end{equation}

\begin{equation}\label{eq:f1}
F1-score = 2 \times \frac{Precision \times Recall}{Precision + Recall}
\end{equation}

\subsection{Scenarios} \label{Scenarios}
We implement three scenarios for evaluating the performance of the Ensemble Learning-Based Detector compared to the Single Learning-Based Detector.
\subsubsection{Scenario 1 – Evaluation on performance of detector and target models} This scenario aims to assess the capability of detecting malware of deployed detectors as well as target model, which are single or ensemble algorithms.

\subsubsection{Scenario 2 – Targeted attack} In this scenario, the Discriminator in FeaGAN has a similar algorithm (single or ensemble) to the targeted models. In other words, our model is trained to generate adversarial samples to fool the Discriminator or the known target models.

\subsubsection{Scenario 3 – Transfer attack} This is the case where FeaGAN and Target Model have different underlying algorithms. Our goal is to evaluate whether the malicious patterns generated by an ensemble learning algorithm remain evading ability against target detectors based on single algorithms and other ensemble algorithms.

\subsubsection{Scenario 4 – Evaluation on the capability of malware preservation}
Preserving the integrity of mutant samples is a critical criterion to evaluate the performance of the generation method, which can be categorized into 3 aspects.

\begin{itemize}
    \item \textit{Format-preserving}: All mutant malware samples need to be in the same PE format as the original ones. To verify this, we use a tool called \textbf{\textit{file}} installed in Cuckoo VM.
    \item  \textit{Executability-preserving}: We aim to check where mutants are executable by running them in Cuckoo Sandbox. Then, we save information about the signatures of samples for evaluating executability-preserving, as well as later maliciousness-preserving. The mutant sample is considered executability-preserving when generating at least one dynamic signature (generating network traffic, accessing REGISTRY or files on the system, etc.).
    \item  \textit{Maliciousness-preserving}: This evaluation is indicated by the number of similar signatures in the original malware and its mutant sample. So, the maliciousness is claimed to be preserved if the difference is less than a predefined threshold, which is 6 signatures in our study.
\end{itemize}

\subsection{Experimental results and analysis}

\subsubsection{Scenario 1}

\paragraph{Performance of malware detectors}
    
The performance of malware detectors in our proposed FeaGAN as well as target models are presented in \textbf{Table~\ref{table:malware_FeaGAN_performance}}. 

As we can see, among single algorithms, LR achieves the best performance in most metrics. Besides, LR, KNN and DT are the best three ones with all metrics above 89.5\%. 

\begin{table*}[!t]
\centering
\caption{Performance of Target Model in DQEAF}
\begin{tabular}{ccccccc}
\hline
\multicolumn{2}{c}{\textbf{Target Model}}                                                     & \textbf{AUC}     & \textbf{Accuracy} & \textbf{Precision} & \textbf{Recall}  & \textbf{F1-Score} \\ \hline
\multicolumn{1}{c}{\multirow{5}{*}{\textbf{Single}}}        & LR             & 0.66738 & 0.75660  & 0.89648   & 0.58020 & 0.70447  \\ \cline{2-7} 
\multicolumn{1}{c}{}                               & DT                   & \cellcolor{blue!25}\textbf{0.97195} & \cellcolor{blue!25}\textbf{0.97195}  & \cellcolor{blue!25}\textbf{0.96705}   & \cellcolor{blue!25}\textbf{0.97720} & \cellcolor{blue!25}\textbf{0.97210}  \\ \cline{2-7} 
\multicolumn{1}{c}{}                               & Naive Bayes                     & 0.86779 & 0.75250  & 0.94236   & 0.53790 & 0.68487  \\ \cline{2-7} 
\multicolumn{1}{c}{}                               & MLP                             & 0.82203 & 0.82185  & 0.93641   & 0.69060 & 0.79494  \\ \cline{2-7} 
\multicolumn{1}{c}{}                               & KNN                      & 0.81440 & 0.76705  & 0.71321   & 0.89330 & 0.79316  \\ \hline
\multicolumn{1}{c}{\multirow{5}{*}{\textbf{Homogenenous}}}  & RF                   & \cellcolor{blue!25}\textbf{0.99895} & \cellcolor{blue!25}\textbf{0.99205}  & \cellcolor{blue!25}\textbf{0.98897}   & 0.99520 & \cellcolor{blue!25}\textbf{0.99207}  \\ \cline{2-7} 
\multicolumn{1}{c}{}                               & AdaBoost                        & 0.99522 & 0.97330  & 0.95721   & 0.99060  & 0.97376  \\ \cline{2-7} 
\multicolumn{1}{c}{}                               & GB               & 0.99877 & 0.98450  & 0.97259   & \cellcolor{blue!25}\textbf{0.99710} & 0.98469  \\ \cline{2-7} 
\multicolumn{1}{c}{}                               & Bagging                         & 0.99357 & 0.97870  & 0.97349   & 0.98420 & 0.97882  \\ \cline{2-7} 
\multicolumn{1}{c}{}                               & GB (gym-malware) & 0.99079 & 0.95200    & 0.96217   & 0.94100   & 0.95147  \\ \hline
\multicolumn{1}{c}{\multirow{3}{*}{\textbf{Heterogeneous}}} & Voting                          & 0.96844 & 0.84550  & 0.96401   & 0.71780 & 0.82288  \\ \cline{2-7} 
\multicolumn{1}{c}{}                               & Stacking(5, DT)           & 0.92471 & 0.92355  & 0.91890   & 0.92910 & 0.92397  \\ \cline{2-7} 
\multicolumn{1}{c}{}                               & Stacking(5, RF)           &\cellcolor{blue!25}\textbf{0.99757} & \cellcolor{blue!25}\textbf{0.98775}  & \cellcolor{blue!25}\textbf{0.97730}   & \cellcolor{blue!25}\textbf{0.99870} & \cellcolor{blue!25}\textbf{0.98788}  \\ \hline
\end{tabular}%
\label{table:target_model_performance}
\end{table*}
In the case of homogeneous ensemble learning, RF seems to be better in performance than other algorithms. However, GB or Bagging still have their strength in some other metrics.

In the heterogeneous ensemble learning, besides voting, we have stacking cases indicated by \textit{Stacking(n, Algo)}, where \textit{n} and \textit{Algo} are the number of base estimators and the algorithm of final estimator, respectively. For example, \textit{Stacking(3, LR)} represents Stacking ensemble learning using 3 base estimators and LR as the final estimator. Note that, algorithms in the 3-estimator stacking case are the 3 best single algorithms, which are LR, KNN and DT as mentioned before.
Via the results in \textbf{Table~\ref{table:malware_FeaGAN_performance}}, it seems unable to figure out the best model in the heterogeneous approach. Moreover, a change in the number of estimators does not result in a clear increase or decrease trend in performance of all models. The only notable case is that KNN has most of its metrics dropped significantly when having more base estimators, except Recall reaching the best value among all heterogeneous algorithms.
This can be caused by the mismatch between the algorithm and the data set, as well as the lack of data to provide reliable statistics. However, it can be noted that the Stack technique outperforms the Voting one. Besides, in general, ensemble algorithms achieve a little better performance compared to their single ones.


\paragraph{Performance of target models}

\textbf{Table~\ref{table:target_model_performance}} offers the malware-detecting capability of target models in the RL model. Note that, due to the slight difference when changing the number of base estimators, stacking-based models are designed with 5 base estimators. Moreover, the two cases of stacking differ in the used base and final estimators. The first case utilizes single algorithms DT, LR, Naive Bayes, MLP, and KNN for base estimators, and the best of them – DT as the final. In the second case, underperforming single algorithms are substituted by ensemble ones to have DT, RF, Adaboost, Bagging, and GB as base estimators and the final estimator of RF.

Clearly, DT and RF outperform other single and homogeneous algorithm-based models with the best values in most metrics.
Moreover, the experimental results indicate that using ensemble algorithms in the stacking method yields more favorable outcomes. More specifics, the metrics of the second stacking case surpass those of the first stacking approach by 5.8\% to 7.2\%.

Besides, based on the above results, to simplify the comparison in later scenarios, we chose the best algorithms with the highest performance to represent each case of the target models. Specifically, DT and KNN are used as the representatives for single learning detectors. Meanwhile, RF and GB (both ours and gym-malware-based) are examples of Homogeneous Ensemble Learning and Stacking(5, RF) is for Heterogeneous Ensemble Learning.

After training, we also consider the performance of the target models made on testing 2,000 malware samples, as in \textbf{Table~\ref{table:independent_original}}. The results indicate that the target models can identify malware effectively, with low evasion rates of mostly under 10\% and a high average score in VirusTotal.


\begin{table}[!t]
\caption{Ability of malware detectors to identify 2,000 original malicious samples}
\centering
\begin{tabular}{lccc}
\hline
\multirow{2}{*}{\textbf{Target model}}    & \multicolumn{3}{c}{\textbf{Before the mutation}}                                                         \\ \cline{2-4} 
                                 & \multicolumn{1}{c}{\textbf{Detected}} & \multicolumn{1}{c}{\textbf{\makecell{Evasion\\ Rate (ER)}}} & \textbf{\makecell{Avarage\\VirusTotal score}} \\ \hline
\textbf{DT} & \multicolumn{1}{c}{1,945} & \multicolumn{1}{c}{2.75\%} & \multirow{6}{*}{58/73} \\ \cline{1-3}
\textbf{KNN}                       & \multicolumn{1}{c}{1,724}    & \multicolumn{1}{c}{13.80 \%}      &                          \\ \cline{1-3}
\textbf{RF}                     & \multicolumn{1}{c}{1,954}    & \multicolumn{1}{c}{2.30 \%}       &                          \\ \cline{1-3}
\textbf{GB}                 & \multicolumn{1}{c}{1,967}    & \multicolumn{1}{c}{1.65 \%}      &                          \\ \cline{1-3}
\textbf{Stacking(5, RF)}           & \multicolumn{1}{c}{1,983}    & \multicolumn{1}{c}{0.85 \%}      &                          \\ \cline{1-3}
\textbf{\makecell{GB (gym-malware)}} & \multicolumn{1}{c}{1,997}    & \multicolumn{1}{c}{0.15 \%}      &                          \\ \hline
\end{tabular}%

\label{table:independent_original}
\end{table}

\begin{table}[!t]
\centering
\caption{Recall metric on Adversarial features of Malware Detector in FeaGAN}
\begin{tabular}{ccc}
\hline
\multicolumn{2}{c}{\textbf{Algorithms}}                                                        & \textbf{Recall on adversarial features} \\ \hline
\multicolumn{1}{c}{\multirow{5}{*}{\textbf{Single}}}        & Bernoulli               & 0                      \\ \cline{2-3} 
\multicolumn{1}{c}{}                                        & Naive Bayes             & 0                      \\ \cline{2-3} 
\multicolumn{1}{c}{}                                        & DT           & 0                      \\ \cline{2-3} 
\multicolumn{1}{c}{}                                        & LR     & 0                      \\ \cline{2-3} 
\multicolumn{1}{c}{}                                        & KNN              & 0.215                \\ \hline
\multicolumn{1}{c}{\multirow{4}{*}{\textbf{Homogeneous}}}   & RF           & 0                      \\ \cline{2-3} 
\multicolumn{1}{c}{}                                        & Bagging                 & 0.49625                \\ \cline{2-3} 
\multicolumn{1}{c}{}                                        & AdaBoost                & 0.04968                \\ \cline{2-3} 
\multicolumn{1}{c}{}                                        & GB       & 0                      \\ \hline
\multicolumn{1}{c}{\multirow{7}{*}{\textbf{Heterogeneous}}} & Voting                  & 0                      \\ \cline{2-3} 
\multicolumn{1}{c}{}                                        & Stacking(3, LR)     & 0                      \\ \cline{2-3} 
\multicolumn{1}{c}{}                                        & Stacking(3, KNN)   & 0.00625                \\ \cline{2-3} 
\multicolumn{1}{c}{}                                        & Stacking(3, DT) & 0                      \\ \cline{2-3} 
\multicolumn{1}{c}{}                                        & Stacking(5, LR)      & 0                      \\ \cline{2-3} 
\multicolumn{1}{c}{}                                        & Stacking(5, KNN)    & 0                      \\ \cline{2-3} 
\multicolumn{1}{c}{}                                        & Stacking(5, DT)  & 0                      \\ \hline
\end{tabular}%

\label{table:recall_after_FeaGAN}
\end{table}

\begin{table}[!b]
\centering
\caption{The evasion result of 2,000 mutant samples on target models and VirusTotal in targeted attack}
\begin{tabular}{lrrrr}
\hline
\multirow{2}{*}{\textbf{Target model}} & \multicolumn{4}{c}{\textbf{After the mutation}}                                                                     \\ \cline{2-5} 
                                  & \multicolumn{1}{c}{\textbf{Detected}} & \multicolumn{1}{c}{\textbf{\makecell{ER}}} & \textbf{\makecell{Virus-\\Total Score}} & \textbf{\makecell{Increased\\ ER}}\\ \hline
\textbf{DT}                     & \multicolumn{1}{c}{711}            & \multicolumn{1}{c}{64.45\%}         & \multicolumn{1}{c}{40/72}   & \multicolumn{1}{c}{61.7\%}            \\ \hline
\textbf{KNN}                        & \multicolumn{1}{c}{1,694}           & \multicolumn{1}{c}{15.3\%}          & \multicolumn{1}{c}{52/71}  & \multicolumn{1}{c}{1.5\%}               \\ \hline
\textbf{RF}                     & \multicolumn{1}{c}{1,902}           & \multicolumn{1}{c}{4.9\%}           & \multicolumn{1}{c}{48/72}  & \multicolumn{1}{c}{2.6\%}               \\ \hline
\textbf{GB}                 & \multicolumn{1}{c}{1,953}           & \multicolumn{1}{c}{2.35\%}          & \multicolumn{1}{c}{46/71}                 & \multicolumn{1}{c}{0.7\%}\\ \hline
\textbf{Stacking(5, RF)}             & \multicolumn{1}{c}{1,520}           & \multicolumn{1}{c}{24\%}            & \multicolumn{1}{c}{53/71}     & \multicolumn{1}{c}{23.15\%}            \\ \hline
\textbf{GB (gym-malware)}   & \multicolumn{1}{c}{1,837}           & \multicolumn{1}{c}{8.15\%}            & \multicolumn{1}{c}{56/73}            & \multicolumn{1}{c}{8\%}     \\ \hline
\end{tabular}%

\label{table:scenario_1}
\end{table}

\begin{figure*}[!b]
    \centering
    \includegraphics[width=0.7\textwidth]{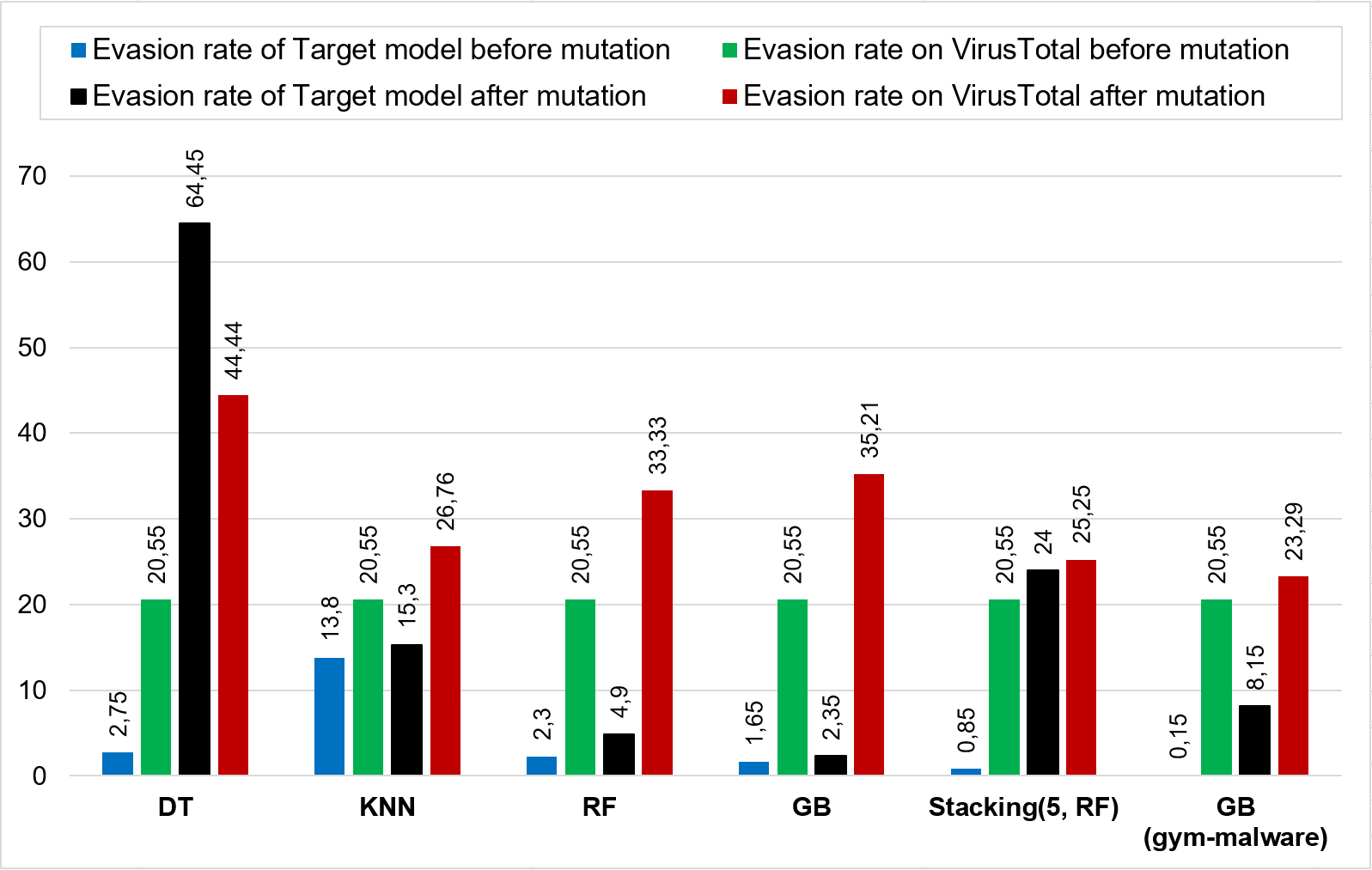}
    \caption{The evasion rate of mutant samples.}
    \label{Bieudo:kichban1}
\end{figure*}

\begin{table*}[!t]
\centering
\caption{Ability to detect mutant malware samples when performing a transfer attack (\textbf{total:} 2,000 samples)}
\resizebox{\textwidth}{!}{%
\begin{tabular}{ccccccccccccc}
\hline
 &
  \multicolumn{12}{c}{\textbf{Target model}} \\ \cline{2-13} 
 &
  \multicolumn{2}{c}{\textbf{DT}} &
  \multicolumn{2}{c}{\textbf{KNN}} &
  \multicolumn{2}{c}{\textbf{RF}} &
  \multicolumn{2}{c}{\textbf{GB}} &
  \multicolumn{2}{c}{\textbf{Stacking(5, RF)}} &
  \multicolumn{2}{c}{\textbf{GB (gym-malware)}} \\ \cline{2-13} 
\multirow{-3}{*}{\textbf{Mutant sample by}} &
  \textbf{Detected} &
  \textbf{ER} &
  \textbf{Detected} &
  \textbf{ER} &
  \textbf{Detected} &
  \textbf{ER} &
  \textbf{Detected} &
  \textbf{ER} &
  \textbf{Detected} &
  \textbf{ER} &
  \textbf{Detected} &
  \textbf{ER} \\ \hline
\textbf{DT-M} &
  \cellcolor[HTML]{FFFFFF}\cellcolor{blue!25}\textbf{711} &
  \cellcolor[HTML]{FFFFFF}\cellcolor{blue!25}\textbf{64.45} &
  \cellcolor[HTML]{FFFFFF}1,614 &
  \cellcolor[HTML]{FFFFFF}19.3 &
  \cellcolor[HTML]{FFFFFF}1,391 &
  \cellcolor[HTML]{FFFFFF}30.45 &
  \cellcolor[HTML]{FFFFFF}\cellcolor{blue!25}\textbf{1,915} &
  \cellcolor[HTML]{FFFFFF}\cellcolor{blue!25}\textbf{4.25} &
  \cellcolor[HTML]{FFFFFF}1,915 &
  \cellcolor[HTML]{FFFFFF}4.25 &
  \cellcolor[HTML]{FFFFFF}1,965 &
  \cellcolor[HTML]{FFFFFF}1.75 \\ \hline
\textbf{KNN-M} &
  \cellcolor[HTML]{FFFFFF}1,889 &
  \cellcolor[HTML]{FFFFFF}5.55 &
  \cellcolor[HTML]{FFFFFF}1,694 &
  \cellcolor[HTML]{FFFFFF}15.3 &
  \cellcolor[HTML]{FFFFFF}1,902 &
  \cellcolor[HTML]{FFFFFF}4.9 &
  \cellcolor[HTML]{FFFFFF}1,957 &
  \cellcolor[HTML]{FFFFFF}2.15 &
  \cellcolor[HTML]{FFFFFF}1,965 &
  \cellcolor[HTML]{FFFFFF}1.75 &
  \cellcolor[HTML]{FFFFFF}\cellcolor{blue!25}\textbf{1,836} &
  \cellcolor[HTML]{FFFFFF}\cellcolor{blue!25}\textbf{8.2} \\ \hline
\textbf{RF-M} &
  \cellcolor[HTML]{FFFFFF}1,905 &
  \cellcolor[HTML]{FFFFFF}4.75 &
  \cellcolor[HTML]{FFFFFF}1,692 &
  \cellcolor[HTML]{FFFFFF}15.4 &
  \cellcolor[HTML]{FFFFFF}1,902 &
  \cellcolor[HTML]{FFFFFF}4.9 &
  \cellcolor[HTML]{FFFFFF}1957 &
  \cellcolor[HTML]{FFFFFF}2.15 &
  \cellcolor[HTML]{FFFFFF}1,964 &
  \cellcolor[HTML]{FFFFFF}1.8 &
  \cellcolor[HTML]{FFFFFF}1,837 &
  \cellcolor[HTML]{FFFFFF}8.15 \\ \hline
\textbf{GB-M} &
  \cellcolor[HTML]{FFFFFF}1,910 &
  \cellcolor[HTML]{FFFFFF}4.5 &
  \cellcolor[HTML]{FFFFFF}1,697 &
  \cellcolor[HTML]{FFFFFF}15.15 &
  \cellcolor[HTML]{FFFFFF}1,908 &
  \cellcolor[HTML]{FFFFFF}4.6 &
  \cellcolor[HTML]{FFFFFF}1,953 &
  \cellcolor[HTML]{FFFFFF}2.35 &
  \cellcolor[HTML]{FFFFFF}1,961 &
  \cellcolor[HTML]{FFFFFF}1.95 &
  \cellcolor[HTML]{FFFFFF}1,842 &
  \cellcolor[HTML]{FFFFFF}7.9 \\ \hline
\textbf{Stacking(5,   RF)-M} &
  \cellcolor[HTML]{FFFFFF}912 &
  \cellcolor[HTML]{FFFFFF}54.4 &
  \cellcolor[HTML]{FFFFFF}\cellcolor{blue!25}\textbf{1,559} &
  \cellcolor[HTML]{FFFFFF}\cellcolor{blue!25}\textbf{22.05} &
  \cellcolor[HTML]{FFFFFF}\cellcolor{blue!25}\textbf{1,041} &
  \cellcolor[HTML]{FFFFFF}\cellcolor{blue!25}\textbf{47.95} &
  \cellcolor[HTML]{FFFFFF}1,961 &
  \cellcolor[HTML]{FFFFFF}1.95 &
  \cellcolor[HTML]{FFFFFF}\cellcolor{blue!25}\textbf{1,520} &
  \cellcolor[HTML]{FFFFFF}\cellcolor{blue!25}\textbf{24} &
  \cellcolor[HTML]{FFFFFF}1,954 &
  \cellcolor[HTML]{FFFFFF}2.3 \\ \hline
\textbf{GB (gym-malware)-M} &
  \cellcolor[HTML]{FFFFFF}1,905 &
  \cellcolor[HTML]{FFFFFF}4.75 &
  \cellcolor[HTML]{FFFFFF}1,692 &
  \cellcolor[HTML]{FFFFFF}15.4 &
  \cellcolor[HTML]{FFFFFF}1,902 &
  \cellcolor[HTML]{FFFFFF}4.9 &
  \cellcolor[HTML]{FFFFFF}1,957 &
  \cellcolor[HTML]{FFFFFF}2.15 &
  \cellcolor[HTML]{FFFFFF}1,964 &
  \cellcolor[HTML]{FFFFFF}1.8 &
  \cellcolor[HTML]{FFFFFF}1,837 &
  \cellcolor[HTML]{FFFFFF}8.15 \\ \hline
\end{tabular}%
}
\label{table:scenario_2}
\end{table*}

\subsubsection{Scenario 2}
Initially, we used FeaGAN to create adversarial features.  As shown in \textbf{Table~\ref{table:recall_after_FeaGAN}}, most algorithms have Recall significantly decreased to near 0 when dealing with these crafted features. Bagging and KNN seem to retain a capability to detect adversarial features compared to other algorithms, in which Bagging is the best one.


Subsequently, the result adversarial features are used as raw information for training DQEAF to build complete adversarial samples. Then, those samples are used to perform a targeted attack on target models which are similar to the model in the training phase. The evasion rate of crafted malware on 5 representative models is given in \textbf{Table~\ref{table:scenario_1}}. The higher the evasion rate is, the more effective adversarial samples are. Clearly, DT has obtained the highest ER with 64.45\%. Moreover, compared to the performance in the first scenario of evaluating the target model, our system succeeds in fooling those models with the increased evasion rate. The most effective case is also the DT model, where the evasion rate climbs 61.7\%. Meanwhile, using GB creates adversarial samples with the least effect, which only causes an increase of 0.7\% in the evasion rate. Moreover, in most cases, regardless of the target model in FeaGAN, the average score in VirusTotal drops in classifying created samples, which indicates many classifier engines fail to detect these malware mutants. A more visualized view of those results is given in \textbf{Fig.~\ref{Bieudo:kichban1}}.

\subsubsection{Scenario 3}
The results of the transfer attack are shown in \textbf{Table~\ref{table:scenario_2}}. In general, regardless of the employed algorithm, which can be single or ensemble models, all target models have their evasion rate increased on mutant samples. This indicates that created samples are still effective in transfer attacks. Among the models used in generating mutants, 
the Stack-based ensemble algorithm produces the best results when causing 3 of 5 target models (using a single or ensemble algorithm) to get their highest evasion rates, of which two are transfer attack cases.

\begin{figure*}[!t]
    \centering
    \includegraphics[width=0.72\textwidth]{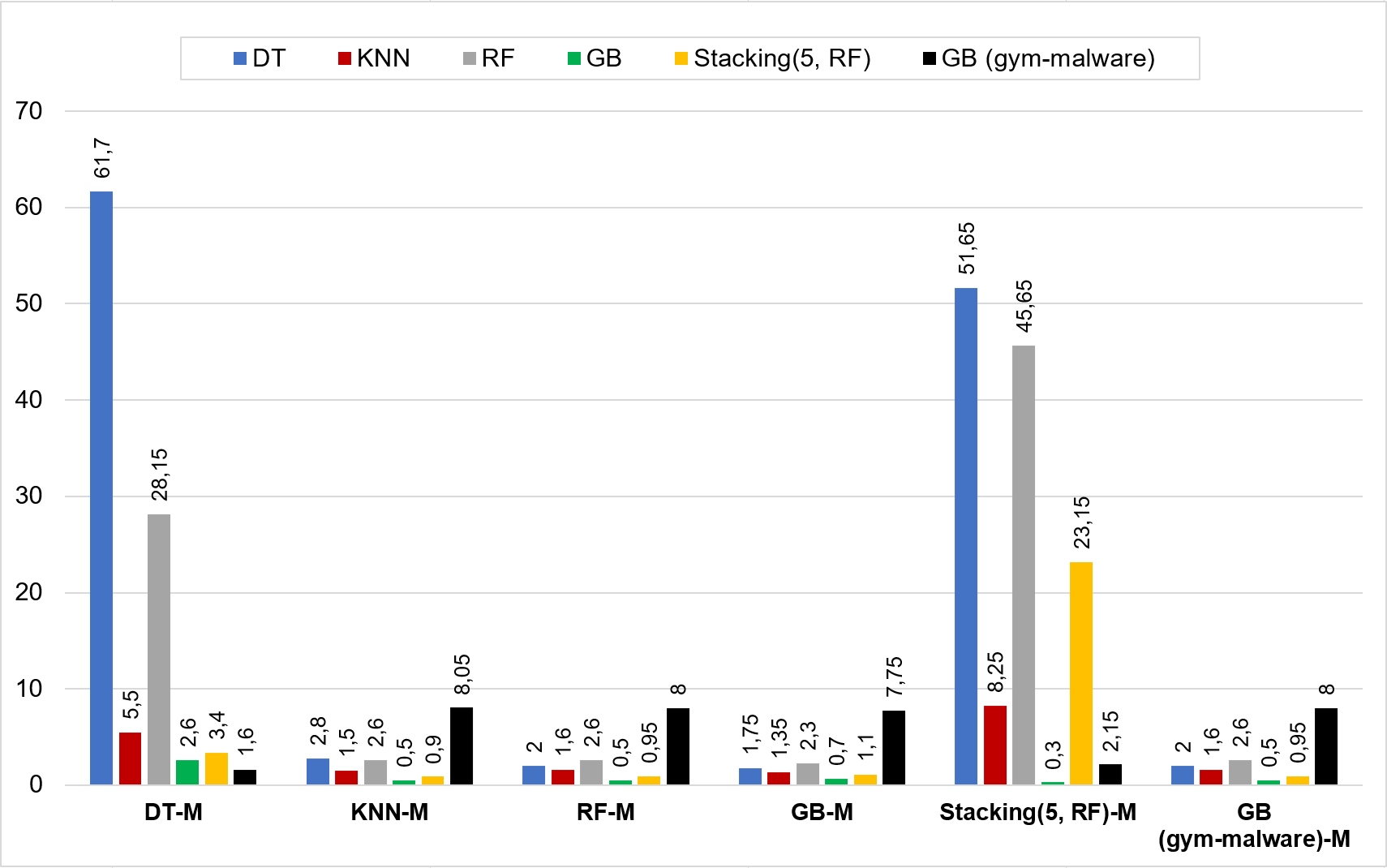}
    \caption{The increased evasion rate of the malware samples after mutation.}
    \label{Bieudo:kichban2}
\end{figure*}

Moreover, to clearly evaluate the effect of mutants on target models, we also consider the difference in ER before and after the transfer attack in terms of increased ER, as in \textbf{Fig.~\ref{Bieudo:kichban2}}. Note that, the models used in making the mutant samples are distinguished from the target models by the suffix “M”. For example, the mutated samples generated by DT algorithms in GAN architecture are denoted by DT-M. Such DT-M samples can be tested in attacks on all ML algorithms including DT, KNN, RF, GB, Stacking(5, RF), and GB (gym-malware). Overall, the evasion rate of all malware samples increased after modification. The highest columns in the group of Stacking(5, RF)-M proves its effectiveness in crafting mutants to fool detectors.

\subsubsection{Scenario 4}
In this scenario, we randomly select 100 result samples from each model used in generating mutants to verify the preservation requirements for the file format. The results, presented in \textbf{Table \ref{table:all_chall_result}}, indicate that 100\% mutant samples guarantee format preservation. This is due to the fact that during mutation, only components that are deemed unrelated are affected, leaving the malicious file intact, thereby ensuring the preservation of the file format.
Moreover, according to our definition of executability-preserving and maliciousness-preserving, the number of mutants meeting those requirements is acceptable. Clearly, it is more problematic to retain the malicious actions than to keep the executability, indicated in the low ratio of maliciousness-preserved mutants. Especially, DT and Stacking(5, RF), which are considered the best algorithms to create the most evasive mutants, seem not to be able to ensure the proper operations of malware with only 7\% and 1\% of truly malicious mutants.

\begin{table}[!]
\centering
\caption{Capability to guarantee the challenges of 100 mutant samples}
\begin{tabular}{cccc}
\hline
\textbf{\makecell{Algorithm\\for mutant}} & \textbf{\makecell{Format\\-preserving}} & \textbf{\makecell{Executability\\-preserving}} & \textbf{\makecell{Maliciousness\\-preserving}} \\ \hline
\textbf{DT}           & 100\% & 60\% & 7\%  \\ \hline
\textbf{KNN}             & 100\% & 94\% & 62\% \\ \hline
\textbf{RF}           & 100\% & 99\% & 63\% \\ \hline
\textbf{GB}       & 100\% & 54\% & 29\% \\ \hline
\textbf{\makecell{Stacking(5, RF)}} & 100\% & 81\% & 1\%  \\ \hline
\textbf{\makecell{GB\\(gym-malware)}}  & 100\% & 94\% & 62\% \\ \hline
\end{tabular}%
\label{table:all_chall_result}
\end{table}
\section{Related work} \label{sec_relatedwork}
There are three challenges of maintaining the semantics of adversarial PE malware for practical and realistic adversarial attacks against PE malware detection that attackers must notice, including format-preserving, executability-preserving, and maliciousness-preserving \cite{P1S_ling2021adversarial}. Unlike images, sounds, or even text, PE Malware must follow the strict formatting rules of a PE file. Therefore, in PE files, transformations in their problem space should be defined within the required format. However, even if we guarantee the format of the file (format-preserving), we cannot keep the executability for PE files (executability-preserving) and the same maliciousness for PE malware (maliciousness-preserving).

\begin{table*}[!b]
\centering
\caption{The summary of related works on malware mutation}
\resizebox{\textwidth}{!}{%
\begin{tabular}{cccccccccc}
\hline
\multirow{6}{*}{\textbf{Year}} &
  \multirow{6}{*}{\textbf{Authors}} &
  \multirow{6}{*}{\textbf{\begin{tabular}[c]{@{}c@{}}Knowledge\\of attacker\end{tabular}}} &
  \multirow{6}{*}{\textbf{\begin{tabular}[c]{@{}c@{}}Manipulation\\Space\end{tabular}}} &
  \multirow{6}{*}{\textbf{Attack Strategy}} &
  \multirow{6}{*}{\textbf{\begin{tabular}[c]{@{}c@{}}PE Malware \\ Detection\end{tabular}}} &
  \multirow{6}{*}{\textbf{Datasets}} &
  \multicolumn{3}{c}{\textbf{Preservation}} \\ \cline{8-10} 
 &
   &
   &
   &
   &
   &
   &
  \rotatebox[origin=c]{90}{\textbf{Format}} &
  \rotatebox[origin=c]{90}{\textbf{Executability}} &
  \rotatebox[origin=c]{90}{\textbf{Maliciousness}} \\ \hline
2017 &
  Hu and Tan \cite{1I_hu2017generating} &
  BB &
  FS &
  GAN &
  \begin{tabular}[c]{@{}c@{}}API call list based \\ malware detectors\end{tabular} &
  Self-collected dataset &
  x &
  x &
   \\ \hline
2019 &
  Liu et al. \cite{ATMPA} &
  WB &
  FS &
  FGSM, C\&W &
  \begin{tabular}[c]{@{}c@{}}Visualization-based \\ malware detectors\end{tabular} &
  \begin{tabular}[c]{@{}c@{}}BIG 2015 \\      VirusTotal\\       Self-collected benign dataset\end{tabular} &
   &
   &
   \\ \hline
2019 &
  Kawai et al. \cite{P24L_ImproveMalGAN} &
  BB &
  FS &
  GAN &
  \begin{tabular}[c]{@{}c@{}}API call list based \\ malware detectors\end{tabular} &
  FFRI Dataset 2018 &
  x &
  x &
   \\ \hline
2019 &
  Fang et al. \cite{P25L_DQEAF} &
  BB &
  PS &
  RL &
  \begin{tabular}[c]{@{}c@{}}GB-based \\ malware detector\end{tabular} &
  VirusTotal &
  x &
  x &
  x \\ \hline
2020 &
  Rosenberg et al. \cite{BADGER} &
  BB &
  FS &
  Evolutionary Algorithm &
  \begin{tabular}[c]{@{}c@{}}API call sequence based \\ malware detectors\end{tabular} &
  Self-collected dataset &
  x &
  x &
   \\ \hline
2020 &
  Yuan et al.  \cite{P26_GAPGAN} &
  BB &
  PS &
  GAN &
  MalConv &
  \begin{tabular}[c]{@{}c@{}}VirusTotal\\      BIG 2015\\      Chocolatey\end{tabular} &
  x &
  x &
   \\ \hline
2021 &
  Lucas et al. \cite{Lucas2021_malmakeover} &
  WB &
  PS &
  Gradient-based &
  \begin{tabular}[c]{@{}c@{}}MalConv\\ AvastNet\end{tabular} &
  \begin{tabular}[c]{@{}c@{}}VirusToTal\\  VirusShare\\      BIG 2015\\      Self-collected benign dataset\end{tabular} &
  x &
  x &
  x \\ \hline
2021 &
  Ebrahimi et al. \cite{AMG_VAC} &
  BB &
  PS &
  RL &
  \begin{tabular}[c]{@{}c@{}}EMBER\\  MalConv\end{tabular} &
  VirusTotal &
  x &
  x &
   \\ \hline
2021 &
  Labaca-Castro et al. \cite{AIMED-RL} &
  BB &
  PS &
  RL &
  \begin{tabular}[c]{@{}c@{}}LightGB-based \\ malware detector\end{tabular} &
  Self-collected dataset &
  x &
  x &
   \\ \hline
 &
  Ours &
  BB &
  PS &
  GAN + RL &
  \begin{tabular}[c]{@{}c@{}}Single algorithm-based   \\malware detectors\\      Ensemble algorithm-based\\ malware detectors\end{tabular} &
  \begin{tabular}[c]{@{}c@{}}VirusTotal\\      Seft-collected benign dataset\end{tabular} &
  x &
  x &
  x \\ \hline
\end{tabular}%
}
\label{table:relatedwork_studies}
\end{table*}

To start with, ATMPA\cite{ATMPA}, proposed by Liu et al., is the white-box adversarial attack strategy in the domain of image-based malware classification tasks. In particular, ATMPA converted the malware sample to a binary texture grayscale image before modifying the corresponding adversarial sample with tiny perturbations provided by two existing adversarial attack techniques - FGSM and C\&W. Experiment results indicated that adversarial noise can reach a 100\% successful attack rate for CNN, SVM, and RF-based malware detectors. Furthermore, the rate of transferability of adversarial samples while attacking various malware detectors might reach 88.7\%. However, the created adversarial grayscale image of the malware sample broke the structure of the original malware and hence could not be executed properly, making ATMPA unsuitable for real-world PE malware detection. In other work, Lucas et al. \cite{Lucas2021_malmakeover} presented a novel category of adversarial attacks based on binary diversification techniques that change binary instructions at the fine-grained function level using two types of functionality-preserving modifications (i.e., in-place randomization and code displacement). To guide the transformations implemented for the PE malware under the white-box option, they used a gradient ascent optimization to select the transformation only if it shifts the embeddings in a direction similar to the gradient of the attack loss function relating to its embeddings. It is evident that practically almost all white-box attacks against PE malware detection, such as the one described above, use optimization of gradient-based approaches as their attack methodologies, regardless of the adversary's space for mutating original samples in feature space or problem space. However, due to the problem-feature space dilemma, it is infeasible and impractical to directly utilize gradient-based optimization algorithms to build realistic adversarial PE malware patterns \cite{space_dilemma}.

In comparison to white-box attacks, black-box attacks are more practicable and realistic in the wild because they rely less on the knowledge of attackers about the target malware detector. For instance, Rosenberg et al. \cite{BADGER} introduced BADGER, an end-to-end adversarial attack system comprised of a set of query-efficient black-box attacks designed to misclassify such API call sequence-based malware detectors while minimizing the number of queries. To preserve the original functionality, their attacks were restricted to only inserting API calls that had no impact or an insignificant impact. The authors presented various attacks with and without knowledge of output likelihood scores to tackle the problem of whether and where the API calls should be added. They conducted two types of adversarial attacks, including the score-based attack and the decision-based attack. For the score-based attack, it employed pre-trained SeqGAN initializing to replicate the API call sequences of benign samples to create the API call. Besides that, it also applied the self-adaptive uniform mixing evolutionary method to optimize the insertion site. Otherwise, the decision-based attack relied on randomness to insert the API call in the same position. To improve the query efficiency of the attacks, they injected objects having a maximum budget and then used a logarithmic backtracking mechanism for removing part of the added API calls while maintaining evasion.

Regarding GAN-based attacks, Hu and Tan \cite{1I_hu2017generating} introduced the MalGAN model to generate PE malware samples, which have the ability to bypass the classification of malware detector, based on the list of API call. Specifically, MalGAN assumed that the attacker knows the entire feature space of the target malware detector. The authors built an alternative detector with the same characteristics of the black-box target model. Then, MalGAN initialized a generation module to minimize the malicious probability of adversarial patterns predicted from the alternative detection machine by adding some unnecessary API calls into the original injected object. In other study, Kawai et al. \cite{P24L_ImproveMalGAN} found some issues from a realistic viewpoint and proposed an improved model from MalGAN called Improved-MalGAN. For instance, Improved-MalGAN used various API call lists during the MalGAN and the training processes of black-box detectors, while the original MalGAN trained them with the same dataset. They also mentioned that the generation of adversarial patterns should not be carried out on diverse types of malware as it may impact the avoidance performance. Besides, Yuan et al. presented GAPGAN \cite{P26_GAPGAN}, a byte-level black-box adversarial attack system based on GAN against DL-based abnormal detection. Their system was built to keep its original functionality and had a high success rate with only short-inserted payloads.

For RL-based attacks, according to Ebrahimi et al. \cite{AMG_VAC}, the actor-critic or DQN is typically used in RL-based adversarial attack methods against PE malware detection and has limitations when handling situations with a large combinatorial state space. By using the variational actor-critic, which has been shown to perform at the cutting edge in managing situations with combinatorial large state space, they offered an improved RL-based adversarial attack framework of AMG-VAC based on gym-malware \cite{P23L_gymmalware, Evading_gym_malware_Anderson}. In addition, DQEAF, a different framework presented by Fang et al. \cite{P25L_DQEAF} that employed DQN to evade PE malware detection, was nearly identical to gym-malware in strategy with a few implementations to increase the effectiveness of the model. Labaca-Castro et al. \cite{AIMED-RL} also offered AIMED-RL, an RL-based adversarial attack framework. The primary distinction between AIMED-RL and other RL-based adversarial attacks is that AIMED-RL provides a novel penalization to the reward function to increase the diversity of the mutated transformation sequences while minimizing the corresponding lengths.

In terms of property preservation including the format, executability, and maliciousness, most adversarial attacks against PE malware detector can only retain the format rather than executability or maliciousness \cite{P1S_ling2021adversarial}. Several adversarial attack methods, like ATMPA\cite{ATMPA}, may damage the fixed layout and structure of the PE format, which is required to load and execute the PE file. Furthermore, there are ways that interact with the feature space \cite{ATMPA, BADGER, 1I_hu2017generating, P24L_ImproveMalGAN} without interacting with the problem space like \cite{Lucas2021_malmakeover, P26_GAPGAN, AMG_VAC, P25L_DQEAF, AIMED-RL}, which is highly likely to be in trouble due to the problem-feature space dilemma. However, it is worth noting that several studies on adversarial attacks such as \cite{1I_hu2017generating,ATMPA,P24L_ImproveMalGAN,BADGER,P26_GAPGAN,AMG_VAC,AIMED-RL} have yet to empirically demonstrate whether the generated adversarial PE malware retains the same level of maliciousness as the original PE malware. These studies are reported and summarized in \textbf{Table \ref{table:relatedwork_studies}}.

Furthermore, several authors propose a variety of techniques to improve the effectiveness of adversarial malware variants. There are two techniques for improving the efficiency of the adversarial samples, including using the variety of attack methods and attacking many classifiers \cite{P1L_li2020adversarial}. In the first strategy, they employ the variety of attack methods to disturb and increase the probability of misclassification of classifiers. Tramer et al. \cite{13L_tramer2019adversarial} suggested employing several mutants to attack a classifier. It produces efficient outcomes in evading the classifier. In the second strategy, they attack the variety of classifiers such that adversarial patterns interact with them as much as possible to maximize the likelihood of avoiding patterns directed at them. In another, Liu et al. \cite{14L_liu2016delving} also proposed improving pattern transferability by attacking a group of mixed DL models rather than a single model. Also, Luca et al. \cite{Demetrio2022_aml_malware} offered a framework that can make a step towards a more systematic and scalable attacking approach for ML algorithms. Their framework is composed of two major blocks. First, the practical manipulations must be defined within the provided application-specific constraints. Secondly, the optimizer is utilized to fine-tune them. It mitigates the four challenges of application-specific, semantics-preserving, automatable, fine-tunable that hinder the application of attacks.


The previously mentioned studies have provided motivation and inspiration for the development of an approach to create mutant malware samples that preserve key properties, such as format, executability, and maliciousness. In this work, we propose a framework that combines FeaGAN and DQEAF to evade anti-malware engines, with a focus on the transferability of adversarial malware samples. Fang's research team has demonstrated the effectiveness of the DQEAF framework, effectively addressing all three challenges \cite{P25L_DQEAF}.

To improve the evasion capability further, we employ an ensemble method for the substitute black-box detector. This approach enhances the interaction of adversarial malware samples and increases their ability to evade the ensemble method-based detector. To ensure the executability and maliciousness of the samples, we test them using the Cuckoo sandbox. The proposed framework aims to create mutant malware samples that maintain their maliciousness while evading detection by anti-malware engines. By combining FeaGAN and DQEAF, we can generate adversarial samples that are transferable across different anti-malware engines. The ensemble method for the substitute black-box detector improves the evasion capability of the samples, making them more effective at avoiding detection. The Cuckoo sandbox testing provides assurance that the samples are both executable and malicious.

Overall, the proposed approach offers a promising solution for the creation of mutant malware samples that evade anti-malware engines while maintaining their maliciousness. The use of FeaGAN and DQEAF, along with the ensemble method for the substitute black-box detector, enables the generation of transferable adversarial samples that are effective at avoiding detection. The Cuckoo sandbox testing provides a reliable means of verifying the executability and maliciousness of the generated samples.








\section{Conclusion} \label{sec_conclusion}

The arms race of malware detection and evasion never comes to an end due to their opposite purposes. Malware can be mutated to create adversarial samples to bypass detectors. Investigating malware generation methods can motivate more proper prevention approaches to deal with mutants. In fact, there is a lack of adversarial sample generation methods operating on the problem space to create a real malware entity instead of a malicious feature vector. In this work, the combination of 3 ML methods: GANs, RL, and ensemble learning to generate adversarial malware samples against ensemble learning-based detectors has achieved certain success. Our proposed FeaGAN and DQEAF have been proven to be effective in crafting malware mutants via experiments while preserving the properties of Windows malware. We also evaluate the effect of the chosen single or ensemble learning algorithms on the overall performance in targeting ensemble learning-based detectors.

In the future, we intend to extend the action space for mutating malware samples to diversify the effectiveness of adversarial samples. Also, other RL algorithms will be investigated to take the least number of actions into original samples for crafting the malware mutants. All collected samples will help malware analysts in-depth insight understand and mitigate the wide spread of the metamorphic malware in critical infrastructures.


%




\ifCLASSOPTIONcaptionsoff
  \newpage
\fi



%



%
\bibliographystyle{IEEEtran}
\bibliography{reference}


\begin{IEEEbiography}[{\includegraphics[width=1in,height=1.25in,clip,keepaspectratio]{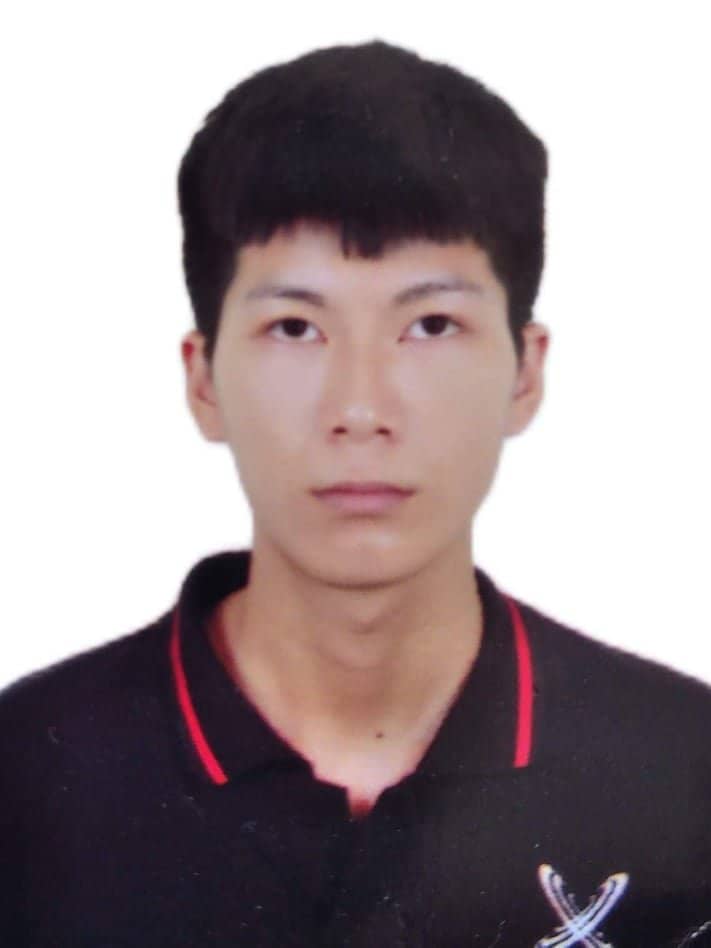}}]{Trong-Nghia To}
received the B. Eng. degree in Information Security from the University of Information Technology, Vietnam National University Ho Chi Minh City (UIT-VNU-HCM) in 2022. From February 2022 until now, he works as a collaborator of the Information Security Laboratory (InSecLab) in UIT. His main research interests include malware analysis, Software-defined networking, blockchain and machine learning.
\end{IEEEbiography}

\begin{IEEEbiography}[{\includegraphics[width=1in,height=1.25in,clip,keepaspectratio]{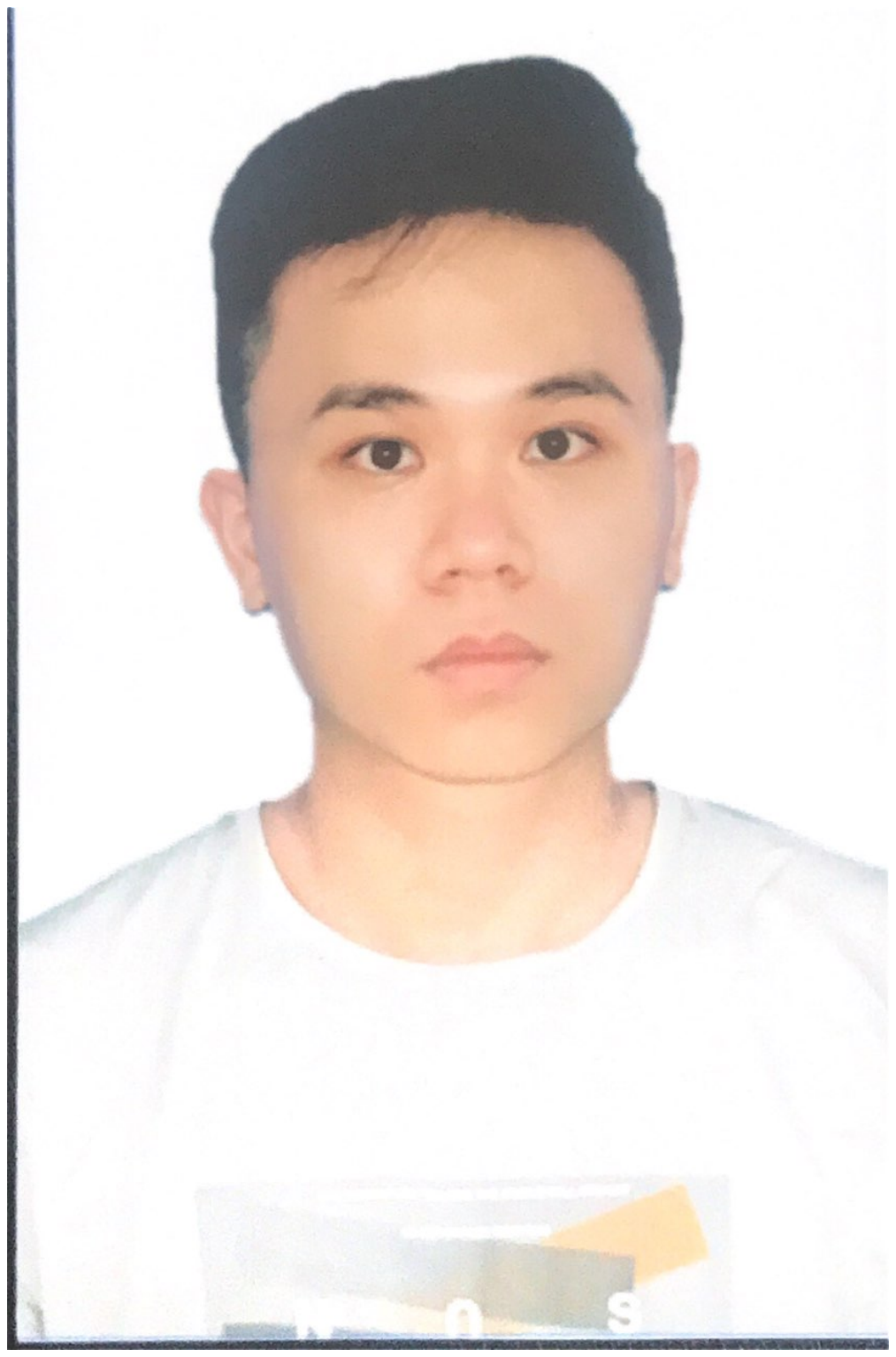}}]{Danh Le Kim}
received the B. Eng. degree in Information Security from the University of Information Technology, Vietnam National University Ho Chi Minh City (UIT-VNU-HCM) in 2022. From February 2022 until now, he works as a collaborator of the Information Security Laboratory (InSecLab) in UIT. His main research interests include malware analysis, digital forensics, pentest and machine learning.
\end{IEEEbiography}




\begin{IEEEbiography}[{\includegraphics[width=1in,height=1.25in,clip,keepaspectratio]{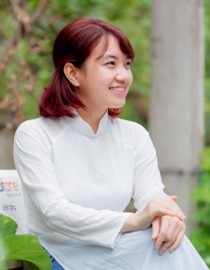}}]{Do Thi Thu Hien}
received the B. Eng. degree in Information Security from the University of Information Technology, Vietnam National University Ho Chi Minh City (UIT-VNU-HCM) in 2017. She received the M.Sc. degree in Information Technology in 2020. From 2017 until now, she works as a member of research group at the Information Security Laboratory (InSecLab) in UIT. Her research interests are Information security \& privacy, Software-defined Networking, and its related security-focused problems.
\end{IEEEbiography}

\begin{IEEEbiography}[{\includegraphics[width=1in,height=1.25in,clip,keepaspectratio]{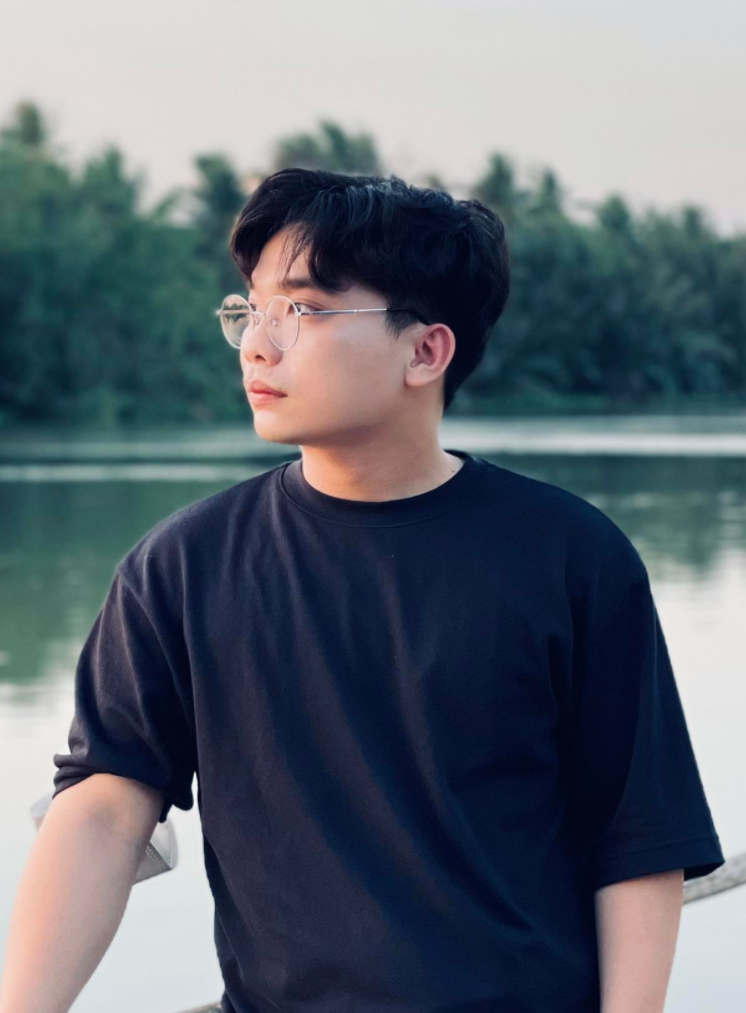}}]{Nghi Hoang Khoa}
received the B.Eng. degree in Information Security from the University of Information Technology, Vietnam National University Ho Chi Minh City (UIT-VNU-HCM) in 2017. He also received the M.Sc. degree in Information Technology in 2022. From 2018 until now, he works as a member of research group at the Information Security Laboratory (InSecLab) in UIT. His main research interests are Information security, malware analysis, Android security, and its related security-focused problems.
\end{IEEEbiography}

\begin{IEEEbiography}[{\includegraphics[width=1in,height=1.25in,clip,keepaspectratio]{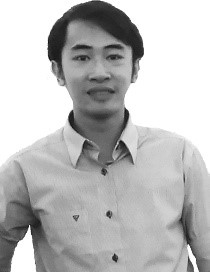}}]{Hien Do Hoang}
received B.E degree in Networking and Communication from the University of Information Technology (UIT), Vietnam National University Ho Chi Minh City (VNU-HCM), Vietnam in 2017. From 2017 to 2018, he worked for a security and network company. He also received the M.Sc. degree in Information Technology in 2020. Currently, he works as a researcher member in Information Security Lab (InSecLab) in UIT, VNU-HCM. His research interests are Software-defined Networking, system and network security, cloud computing and blockchain.

\pdfoutput=1
\end{IEEEbiography}

\begin{IEEEbiography}[{\includegraphics[width=1in,height=1.25in,keepaspectratio]{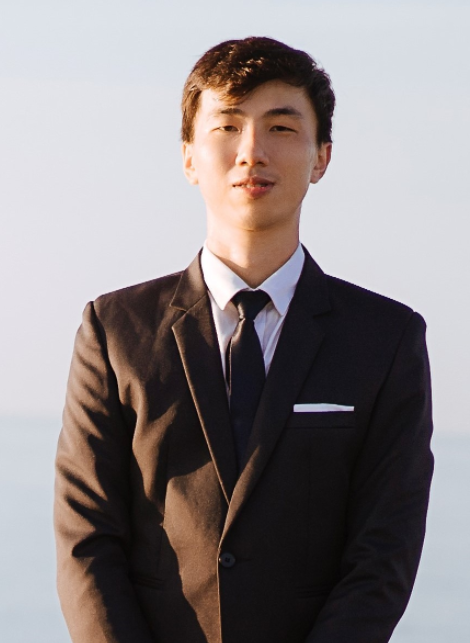}}]{Phan The Duy}
received the B.Eng. and M.Sc. degrees in Software Engineering and Information Technology, respectively from the University of Information Technology (UIT), Vietnam National University Ho Chi Minh City (VNU-HCM), Hochiminh City, Vietnam in 2013 and 2016, respectively. Currently, he is pursuing a Ph.D. degree majoring in Information Technology, specialized in Cybersecurity at UIT, Hochiminh City, Vietnam. He also works as a researcher member in Information Security Laboratory (InSecLab), UIT-VNU-HCM after 5 years in the industry, where he devised several security-enhanced and large-scale teleconference systems. His research interests include Information Security \& Privacy, Software-Defined Networking, Malware Detection and Prevention, Software Security, Digital Forensics, Adversarial Machine Learning, Private Machine Learning, Machine Learning-based Cybersecurity, and Blockchain.
\end{IEEEbiography}

\begin{IEEEbiography}[{\includegraphics[width=1in,height=1.25in,clip,keepaspectratio]{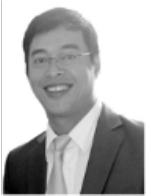}}]{Van-Hau Pham}
obtained his bachelor’s degree in computer science from the University of Natural Sciences of Hochiminh City in 1998. He pursued his master’s degree in Computer Science from the Institut de la Francophonie pour l’Informatique (IFI) in Vietnam from 2002 to 2004. Then he did his internship and worked as a full-time research engineer in France for 2 years. He then persuaded his Ph.D. thesis on network security under the direction of Professor Marc Dacier from 2005 to 2009. He is now a lecturer at the University of Information Technology, Vietnam National University Ho Chi Minh City (UIT-VNU-HCM), Hochiminh City, Vietnam. His main research interests include network security, system security, mobile security, and cloud computing.
\end{IEEEbiography}

\end{document}